\documentclass[iop,twocolappendix,numberedappendix]{emulateapj}
\usepackage{amsmath}
\usepackage{txfonts}
\usepackage{natbib}
\usepackage{graphicx}
\usepackage{color} 

\usepackage{hyperref}

\long\def\dump#1{}

\begin{document}

\title{Flavor-dependent neutrino angular distribution in core-collapse supernovae}
\shorttitle{Neutrino angular distribution}

\preprint{MPP-2017-9, INT-PUB-17-007}

\shortauthors{Tamborra et al.}

\author{Irene Tamborra\altaffilmark{1},
 Lorenz H\"udepohl\altaffilmark{2}, 
 Georg G.~Raffelt\altaffilmark{3},
 and
 Hans-Thomas Janka\altaffilmark{4}
 }

\affil{\altaffilmark{1} Niels Bohr International Academy, Niels Bohr Institute,
Blegdamsvej 17, 2100 Copenhagen, Denmark \\
\altaffilmark{2} Max Planck Computing and Data Facility (MPCDF), Gie{\ss}enbachstr.~2,
85748 Garching, Germany \\
       \altaffilmark{3} Max-Planck-Institut f\"ur Physik
(Werner-Heisenberg-Institut), F\"ohringer Ring 6, 80805 M\"unchen,
Germany\\
       \altaffilmark{4}    Max-Planck-Institut f\"ur Astrophysik,
                        Karl-Schwarzschild-Str.~1, 85748 Garching, Germany}

\begin{abstract}
According to recent studies, the collective flavor evolution of neutrinos
in core-collapse supernovae depends strongly on the flavor-dependent
angular distribution of the local neutrino radiation field, notably on the
angular intensity of the electron-lepton number carried by neutrinos. To
facilitate further investigations of this subject, we study the energy and
angle distributions of the neutrino radiation field computed with
the {\sc Vertex} neutrino-transport code for several spherically symmetric (1D)
supernova simulations (of progenitor masses 11.2, 15 and 25~$M_{\odot}$)
and explain how to extract this information from additional models of the
Garching group. Beginning in the decoupling region (``neutrino sphere''),
the distributions are more and more forward peaked in the radial direction
with an angular spread that is largest for $\nu_e$, smaller for
$\bar\nu_e$, and smallest for $\nu_x$, where $x=\mu$ or $\tau$.  While the
energy-integrated $\nu_e$ minus $\bar\nu_e$ angle distribution has a dip in
the forward direction, it does not turn negative in any of our investigated
cases.
\end{abstract}


\section{Introduction}

Neutrinos are the main agents of energy and lepton-number transport in
core-collapse supernovae (SNe). Within the delayed explosion mechanism of
Bethe and Wilson, neutrinos cause the explosion by shock reheating and
determine nucleosynthesis yields \citep{Janka:2012wk,Burrows:2012ew}.
Neutrinos  will also allow us to monitor core-collapse phenomenology when
high event statistics will be collected by existing and future large-volume
detectors  in the event of a nearby SN explosion \citep{Scholberg:2012id}.

The energies and densities in core-collapse environments are of typical
nuclear-physics scales, i.e., $\mu$ and $\tau$ leptons do not exist except
perhaps for some muons in the hottest regions. Hence, heavy-lepton neutrinos,
often collectively called $\nu_x$, mostly interact by neutral-current
processes, whereas $\nu_e$ and $\bar\nu_e$ interact predominantly by $\beta$
reactions. Neutrino transport, emission, and their detection therefore have a
pronounced flavor dependence.

Developing a reliable theoretical understanding of neutrino flavor evolution
in astrophysical environments with high neutrino density has been
surprisingly difficult because of neutrino-neutrino refraction
\citep{Duan:2006an,Duan:2010bg,Mirizzi:2015eza,Chakraborty:2016yeg}. Flavor
evolution is here a dynamical phenomenon as neutrinos feed back upon
themselves, i.e., one needs to study collective flavor degrees of freedom of
the entire ensemble, not only of individual neutrinos or individual momentum
modes. These collective modes can show instabilities such that flavor
conversion can occur even in regions where the large matter effect suppresses
normal flavor oscillations of individual modes. One generic type of process
is pair conversion $\nu_e\bar\nu_e\leftrightarrow\nu_x\bar\nu_x$ on the
forward-scattering (refractive) level. It does not violate flavor-lepton
number and of course proceeds anyway by normal non-forward scattering, but
can be collectively enhanced even without neutrino flavor mixing.

For the neutrino energies relevant in collapsed stars, the weak processes are
usually described by the four-fermion interaction proportional to the Fermi
constant $G_{\rm F}$. The space-time structure is of current-current form and
implies that the interaction energy between relativistic neutrinos is
proportional to $(1-\cos\vartheta)$, where $\vartheta$ is their relative angle of
propagation. In particular, parallel-moving neutrinos have no mutual
refractive effect at all. Therefore, in this context, we cannot treat
neutrinos as flowing in a purely radial direction.

To account for the crucial neutrino angle distribution, one often
adopted a simple ``bulb model'' of neutrino emission \citep{Duan:2006an}. It
consists of an emitting surface, the ``neutrino sphere,'' as a source of
flavor-dependent neutrino spectra with a common angle distribution. The
latter was often taken to be black-body like, i.e., isotropic into the outer
half-space, or in the form of ``single-angle emission,'' i.e., only one
zenith angle of emission relative to the local radial direction with local
axial symmetry. Some tentative studies used different local angle
distributions between $\nu_x$ and a common one for $\nu_e$ and $\bar\nu_e$ in
a schematic way \citep{Mirizzi:2011tu}. On the other hand, non-trivial angle
distributions, especially between $\nu_e$ and $\bar\nu_e$, may be crucial for
a full appreciation of flavor evolution
\citep{Sawyer:2005jk,Sawyer:2008zs,Sawyer:2015dsa,Chakraborty:2016lct,Dasgupta:2016dbv,Izaguirre:2016gsx,Wu:2017qpc}.

At large distances from the SN core, the neutrino flux is essentially a
narrow ``beam'' with small opening angle. 
In addition, residual scattering
provides a wide-angle ``halo'' which has low intensity, yet  it is responsible for a non-negligible 
contribution to the $(1-\cos\vartheta)$ term \citep{Cherry:2012zw,Sarikas:2012vb}.
By the same token, a ``backward'' flux toward the SN core exists and is
particularly important near the decoupling region where neutrinos flow in all
directions with different intensities. Therefore, a simple treatment of
flavor evolution as a function of radius with only an inner boundary
condition at the arbitrarily defined ``neutrino sphere'' does not necessarily
capture this physical situation \citep{Izaguirre:2016gsx}.  Moreover, if the
$\nu_e$ and $\bar\nu_e$ distributions are sufficiently different, ``fast
flavor instabilities'' can ensue. The latter do not depend on neutrino mass
differences and, specifically, are driven by the angle distribution of the
electron lepton number (ELN) carried by neutrinos.

Despite the role played by neutrino angle distributions in the flavor
evolution, it is difficult to glean enough insight about them from the
published literature on numerical SN simulations. The basic picture of how
neutrinos interact and decouple tells us that the $\nu_e$ distribution should
be broader than that of $\bar\nu_e$, and both broader than that of $\nu_x$.
However, to mimic the local radiation field at some distance $r$ by assuming
these flavors are emitted by different neutrino spheres, i.e., a separate
``bulb model'' for each species, for sure is overly simplistic, especially
close to the SN core where backward propagating neutrinos are important as well.

 The main point of our paper is to fill this gap in the
  literature and to provide a first overview of neutrino energy and
  angle distributions.  In this way we hope to provide crucial input
  information for further studies of collective neutrino flavor
  evolution. In particular, we present angle-dependent distributions
  from hydrodynamical simulations of three SN progenitors with masses
  of 11.2, 15 and $25\,M_\odot$ and characterize their variation as a
  function of the distance from the core. These models were developed
  by \cite{Huedepohl:2013} with the 1D version of the Garching group's
  {\sc Prometheus-Vertex} code.

 In principle, it would be desirable to provide neutrino
  distributions from 3D models. One expects an even richer
  phenomenology and a greater diversity of cases regarding the local
  neutrino distributions. However, the numerical approximations used
  in state-of-the-art 3D simulations
(such as ray-by-ray or two-moment closure schemes)
 are not qualified to provide
  reliable neutrino angle distributions.
Solutions of the Boltzmann transport equation in 3D are needed
  to obtain full phase-space information for the neutrinos. Corresponding
  methods, however, are still being developed. First results for 
  static and stationary conditions in 3D \citep{Sumiyoshi:2014qua} 
  and time-dependent simulations in 2D \citep{Ott:2008jb,Brandt:2010xa,Nagakura:2017mnp}  with 
  multi-angle transport are available, but still suffer from
  major shortcomings, e.g.~the lack of energy-bin coupling
  \citep[]{Ott:2008jb,Brandt:2010xa}, and coarse resolution, especially in the momentum space.
  Therefore, studies similar to the one presented here, but
  for 3D models, will have to wait for the next generation of 
  3D SN simulations including Boltzmann neutrino transport,
  which will require exascale computing.

 In Sec.~\ref{sec:SNmodels} we introduce our models and in
  Sec.~\ref{sec:radial} we describe the angle-integrated features of
  the neutrino field and their variations as functions of radius. In
Sec.~\ref{sec:angles} we study the neutrino distributions as functions
of zenith-angle, radius, and post-bounce time. Conclusions and an
outlook are provided in Sec.~\ref{sec:conclusions}.
Appendix~\ref{sec:RadiationField} reports a glossary of the
definitions usually adopted in neutrino radiative transport vs.\ the
terminology more common in flavor oscillation studies. The neutrino
angle distributions for our models are provided as
\href{http://wwwmpa.mpa-garching.mpg.de/ccsnarchive/data/Tamborra2017/}
     {supplementary material} and instructions on how to use these
     data are provided in Appendix~\ref{sec:nudata}.

\section{Numerical supernova models}\label{sec:SNmodels}
We will explore the characteristics of the neutrino radiation field during
the accretion phase of three spherically symmetric SN simulations with
progenitor masses 11.2, 15 and $25\,M_\odot$. The nuclear equation of state
is from \cite{Lattimer:1991nc} with compressibility modulus $K=220$\,MeV
\citep{Huedepohl:2013}. The simulations were performed with the 1D version 
of the {\sc Prometheus-Vertex} code. It couples an explicit
third-order Riemann-solver-based Newtonian hydrodynamics
code with an implicit three-flavor, multi-energy
group two-moment closure scheme for neutrino transport.
The neutrino transport applied here used three species
$\nu_e$, $\bar\nu_e$ and $\nu_x$ (with $\nu_x=\nu_\mu, \nu_\tau,
\bar{\nu}_\mu, \bar{\nu}_\tau$), and the variable Eddington-factor
closure for the two-moment equations is obtained from a
model Boltzmann equation, whose solution is based on a tangent-ray 
discretization and provides also information 
on the angle-dependent neutrino intensities \citep{RamppJanka2002}, 
which are of central importance for the present work.

General relativistic (GR) corrections are accounted for 
by using an effective gravitational potential 
\citep[case~A of][]{Mareketal2006} and by including
GR redshift and time dilation in the transport.
The relatively small effects of GR ray bending in the NS
environment, however, are ignored in the neutrino treatment,
i.e., the tangent-ray method assumes neutrinos to propagate along
straight paths instead of curved geodesics. 
Tests showed good overall agreement until several 100\,ms after
core bounce \citep{Mareketal2006,Liebendoerferetal2005}
with fully relativistic simulations of the Basel group's
{\sc Agile-Boltztran} code. A more recent comparison with
a GR program \citep{Muelleretal2010}
that combines the {\sc CoCoNuT} hydro solver
\citep{Dimmelmeieretal2002}
with the {\sc Vertex} neutrino transport, revealed
almost perfect agreement except for a few quantities with
deviations of at most 7--10$\%$ until several seconds. 
Our models presented here include the full set of
neutrino reactions described in Appendix~A of 
\cite{Burasetal2006} with the original references given there.
(The simulation setup is analog to the ``full'' opacity
case discussed in \citealt{Huedepohl:2009wh}.)
In particular, we account for
nucleon recoils and thermal motions, nucleon-nucleon
(NN) correlations, weak magnetism, a reduced effective
nucleon mass and quenching of the axial-vector coupling
at high densities, NN bremsstrahlung, $\nu\nu$ scattering, and
neutrino-antineutrino-pair conversions between different
flavors \citep{Burasetal2003}. In addition, we include electron capture
and inelastic neutrino scattering on nuclei 
\citep{Langankeetal2003,Langankeetal2008}.

\begin{figure}
\includegraphics[width=\columnwidth]{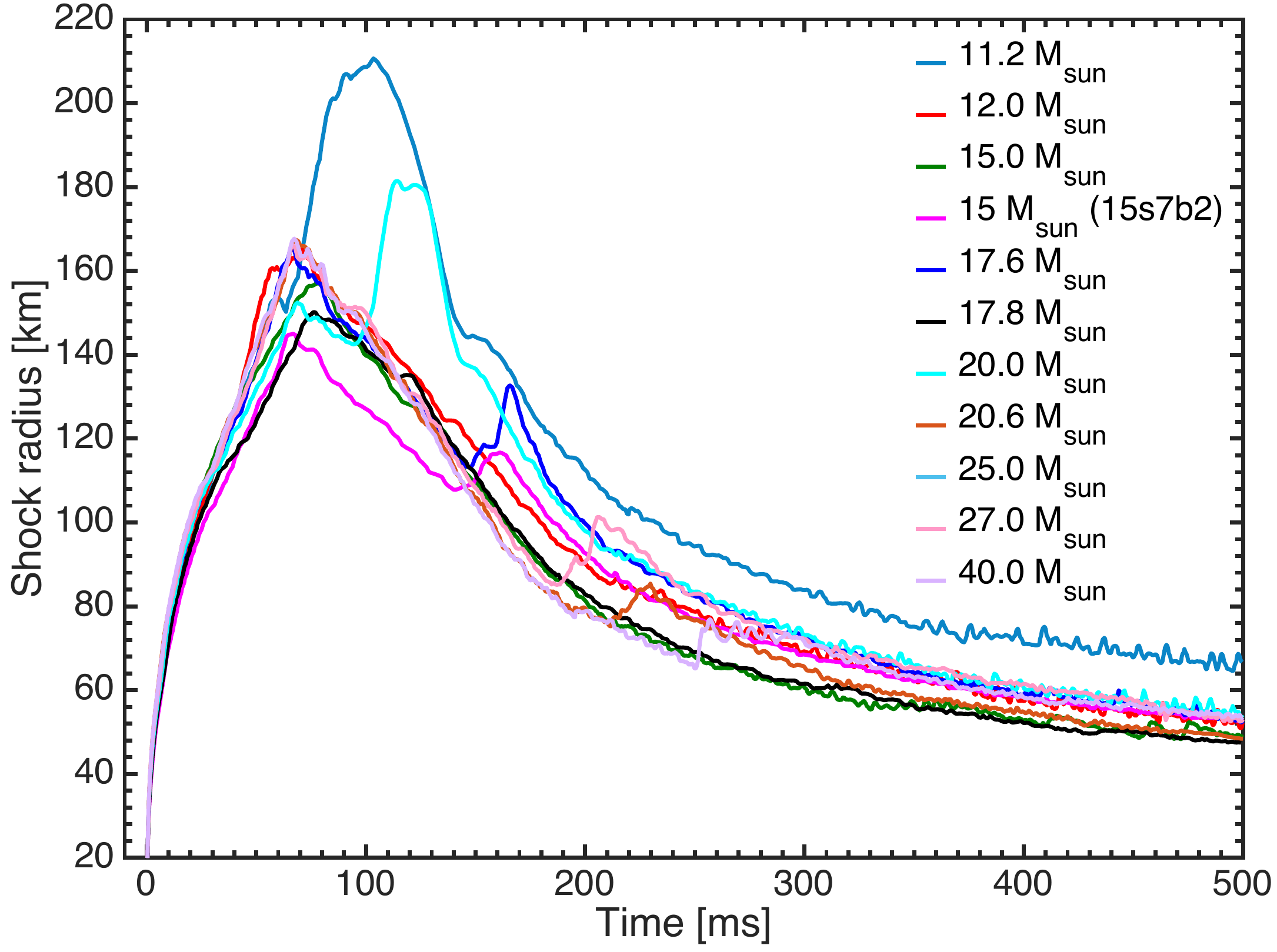}
\caption{Shock radii as functions of post-bounce time for the progenitor
models of \cite{Huedepohl:2013}. The model labelled as ``15s7b2'' 
corresponds to our benchmark case of 15 $M_\odot$.}
\label{fig:shockradii}
\end{figure}

The progenitor models employed for our stellar core-collapse
simulations are taken from \cite{Woosley:1995ip} in the case of the
15\,$M_\odot$ progenitor model s15s7b2 and from \cite{Woosleyetal2002}
for all other investigated cases.
Figure~\ref{fig:shockradii} shows the evolution of the shock radius as a
function of the post-bounce time. All the simulated models exhibit similar
features; see also the angle-integrated neutrino emission properties shown in
Fig.~7 of \cite{Janka:2012sb}. The 11.2 and the $25\,M_\odot$ progenitors
gauge the maximum variation of the neutrino light curves of the broader
simulated mass range of SN progenitors shown in Fig.~7 of
\cite{Janka:2012sb}, while the $15\,M_\odot$ progenitor is an intermediate
case.

For each of the three progenitors discussed here, we inspect the neutrino
angle and
energy distributions at three representative post-bounce times. These are
chosen such as to represent the neutrino-field properties soon after the
shock-breakout burst (early accretion phase), just before the drop in the
luminosity due to the infall of the Si/SiO shell interface, usually occurring
around 200~ms post bounce, and in the late accretion phase. The selected
post-bounce times for each model are $t = 61$, 250 and 550~ms for the
$11.2\,M_\odot$ progenitor, $t = 150$, 280 and 500~ms for $15\,M_\odot$, and
$t = 63$, 250 and 350~ms for $25\,M_\odot$. Unless otherwise specified, the
$t = 280$~ms snapshot of the $15 M_\odot$ model will be used as a benchmark
case.

The fundamental quantity to describe the neutrino radiation field for each
flavor $\nu_\alpha=\nu_e$, $\bar{\nu}_e$ and $\nu_x$ is the spectral
intensity $I_{E,\Omega}$ as explained in Appendix~\ref{sec:RadiationField},
where we provide a glossary of different definitions used in neutrino
radiative transport and in studies of neutrino flavor evolution. The spectral
intensity is the energy of a given species $\nu_\alpha$ that streams at
energy $E$ per unit time through a unit area in a unit solid angle around
direction $\Omega$ perpendicular to the area. It  
has units ${\rm cm}^{-2}~{\rm s}^{-1}~{\rm ster}^{-1}$ (note that the energy
units in numerator and denominator cancel). In spherically
symmetric SN simulations, the intensity is axially symmetric. For each
radial grid point $r_i$, the intensity is discretized at the zenith-angle points
$\mu_{ij}$ (where $\mu=\cos\theta$) on a tangent-ray grid, i.e., 
the zenith-angle grid is
different for different radial positions $r_i$ \citep{RamppJanka2002}. In our
benchmark case, a maximum of 824 angular bins was
used. The energy grid $E_k$ is made
from 21 bins up to 380~MeV with nearly geometric spacing.

The quantity provided by our SN simulations\footnote{The full neutrino data
set presented here is available at the Garching SN archive:
\href{http://wwwmpa.mpa-garching.mpg.de/ccsnarchive/index.html}
{http://wwwmpa.mpa-garching.mpg.de/ccsnarchive/index.html}} is the
monochromatic neutrino intensity for each flavor $\nu_\alpha$, integrated
over the energy bin $[E_{k,\mathrm{min}},E_{k,\mathrm{max}}]$ centered on $E_k$
\begin{equation}
I_{ijk}=\int_{E_{k,\mathrm{min}}}^{E_{k,\mathrm{max}}}I_{E,\Omega}(r_{i},\mu_{ij},E)\, dE\,
\label{eq:monoint}
\end{equation}
in units of ${\rm MeV}~{\rm cm}^{-2}~{\rm s}^{-1}~{\rm ster}^{-1}$.
 Notice that the discrete variables $r_i$ and $E_k$
are at the center of their respective bins. In the following, we will
define the corresponding discrete quantities for the neutrino field as
directly connected to the numerical grid of the simulation.

Traditionally, SN neutrino flavor oscillation studies were concerned with
relatively large distances and typically used a description of the neutrino
field as seen by a distant observer. On the other hand, if we consider the
neutrino-matter
decoupling region, it may be more appropriate to use Lagrangian coordinates
comoving with the local matter flow. The output quantities of the Garching
simulations are given in this comoving frame and we will use it here as well.

\begin{figure*}
\begin{center}
\includegraphics[width=0.95\columnwidth]{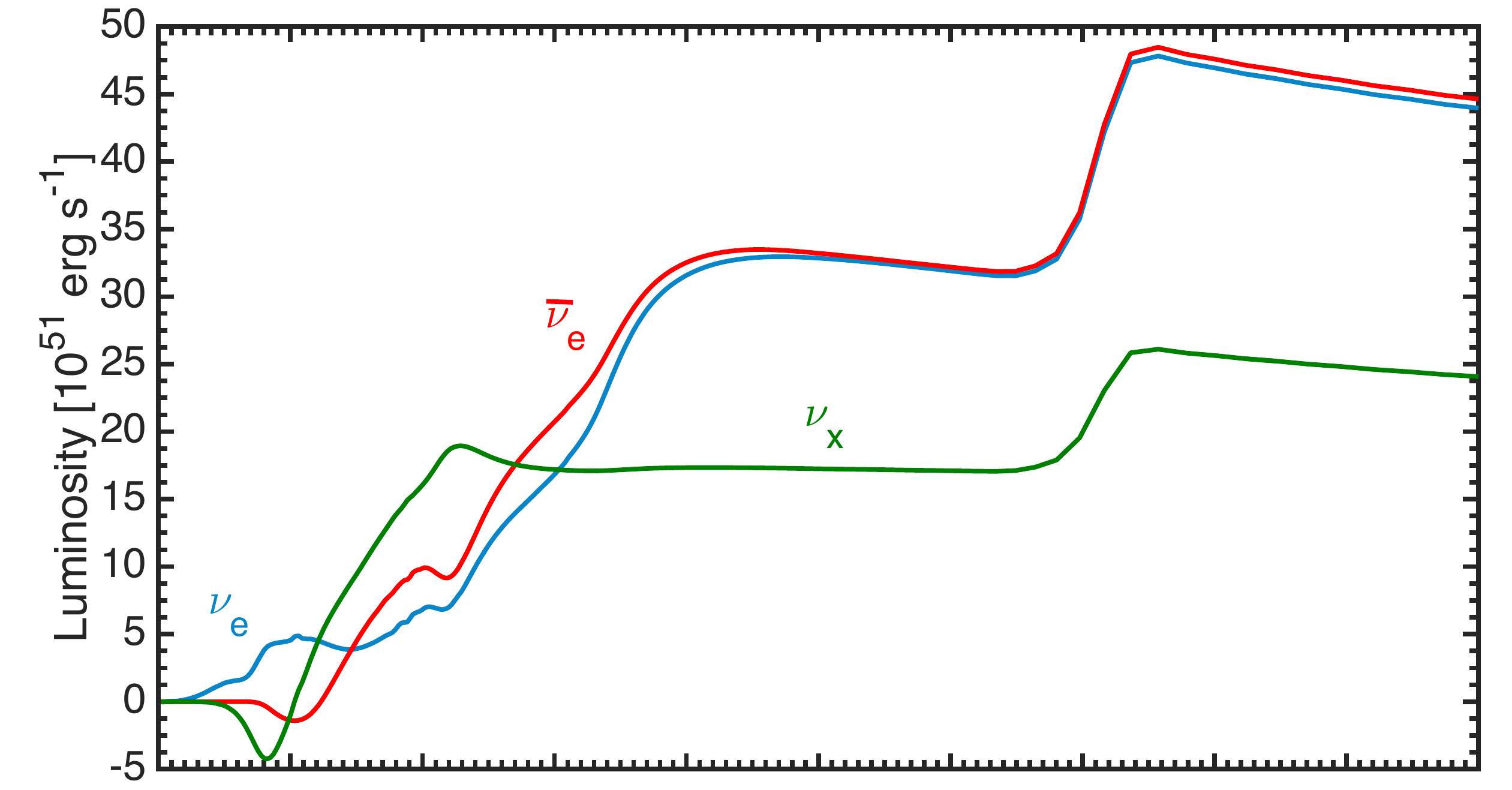}\kern1em\includegraphics[width=0.95\columnwidth]{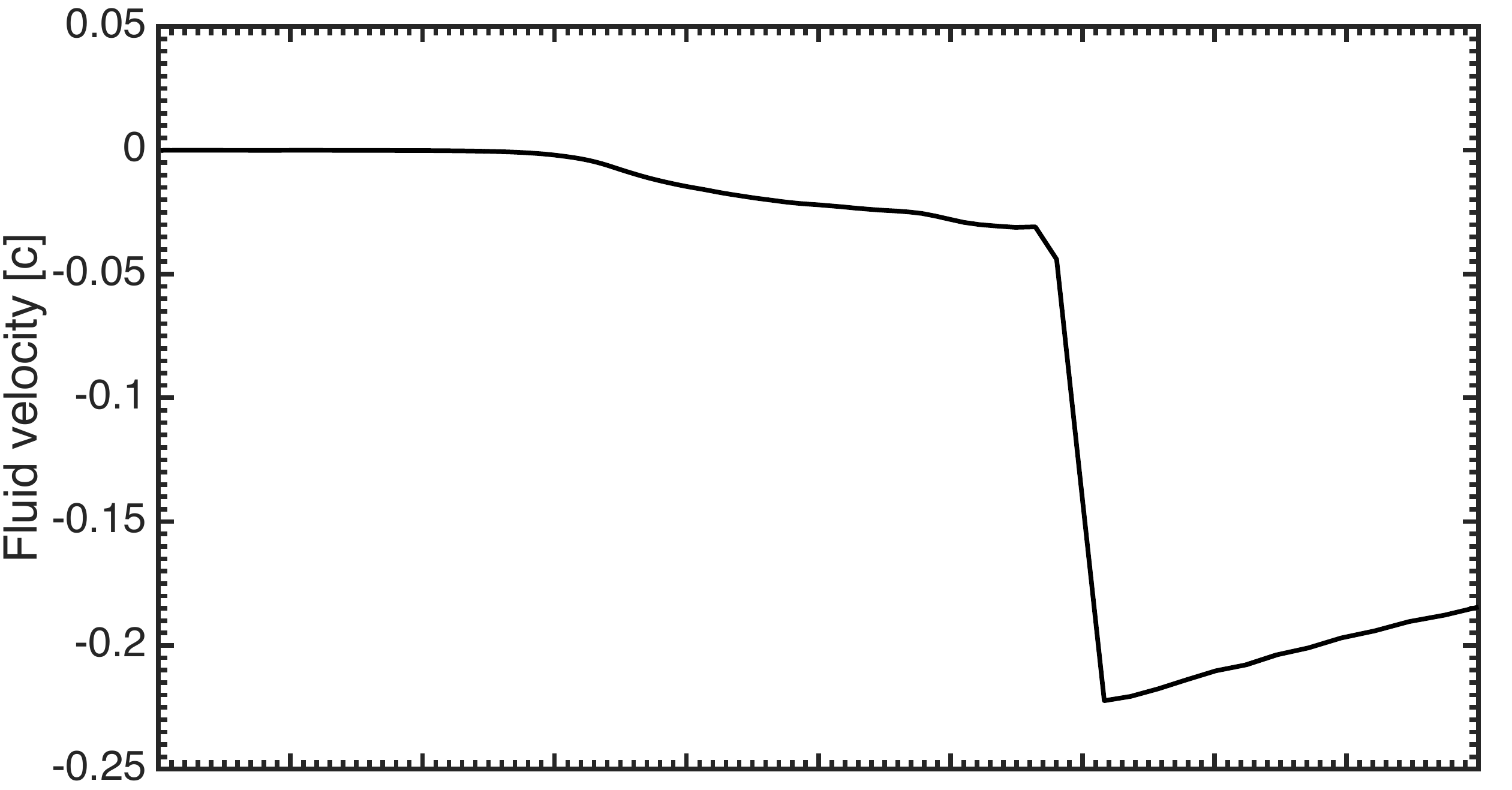}\\
\includegraphics[width=0.95\columnwidth]{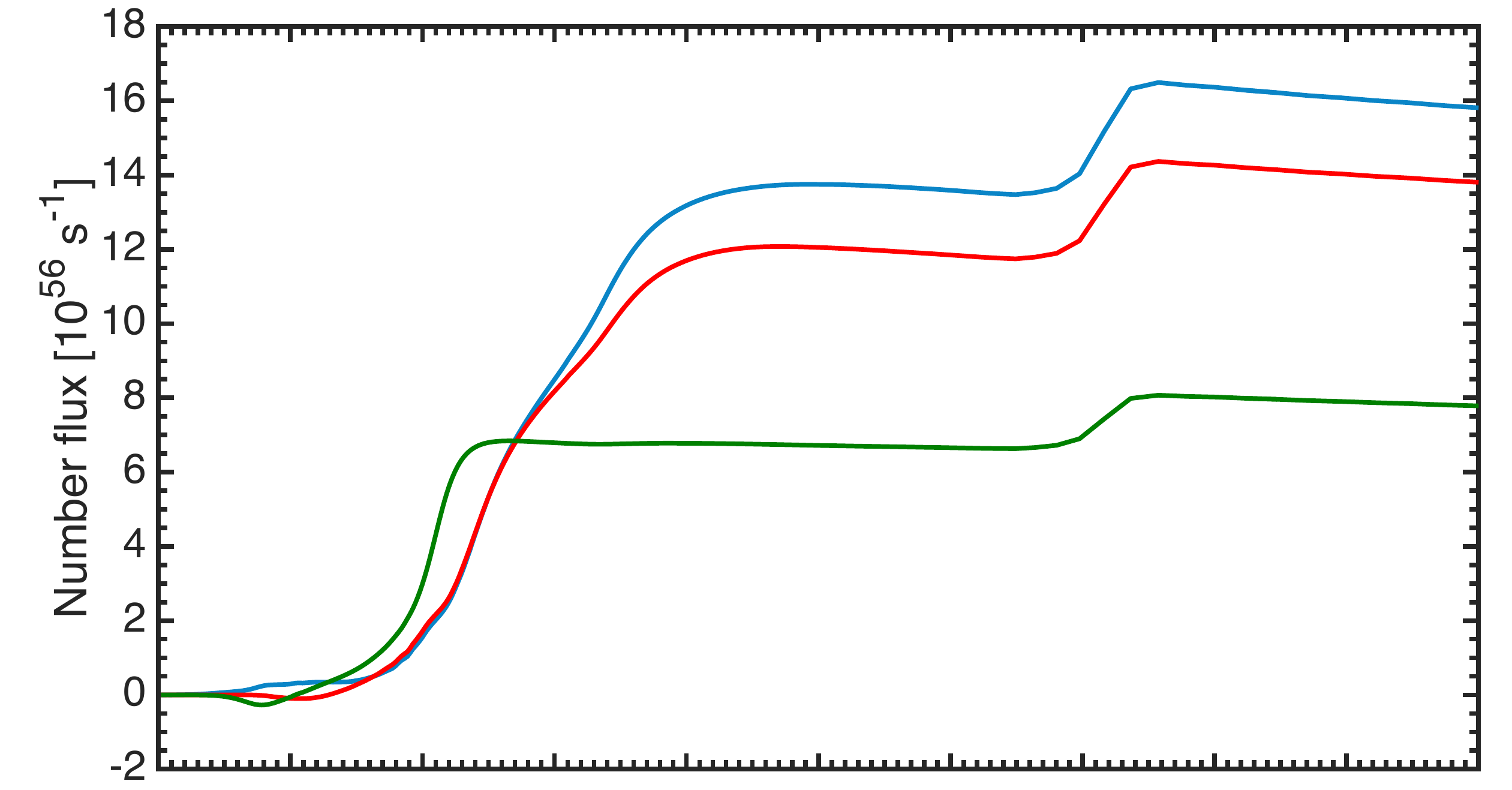}\kern1em\includegraphics[width=0.95\columnwidth]{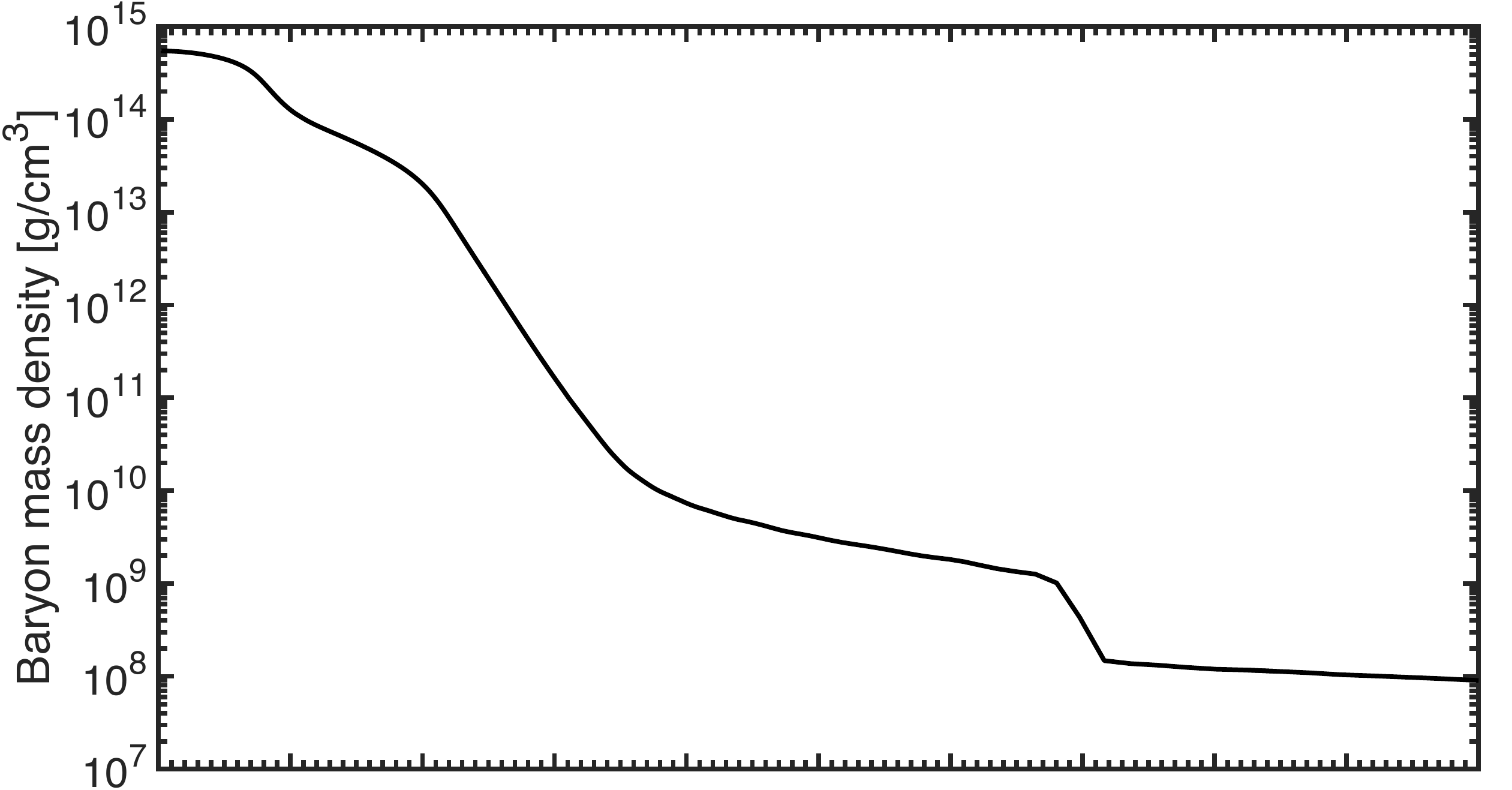}\\
\hspace{1mm}\includegraphics[width=0.94\columnwidth]{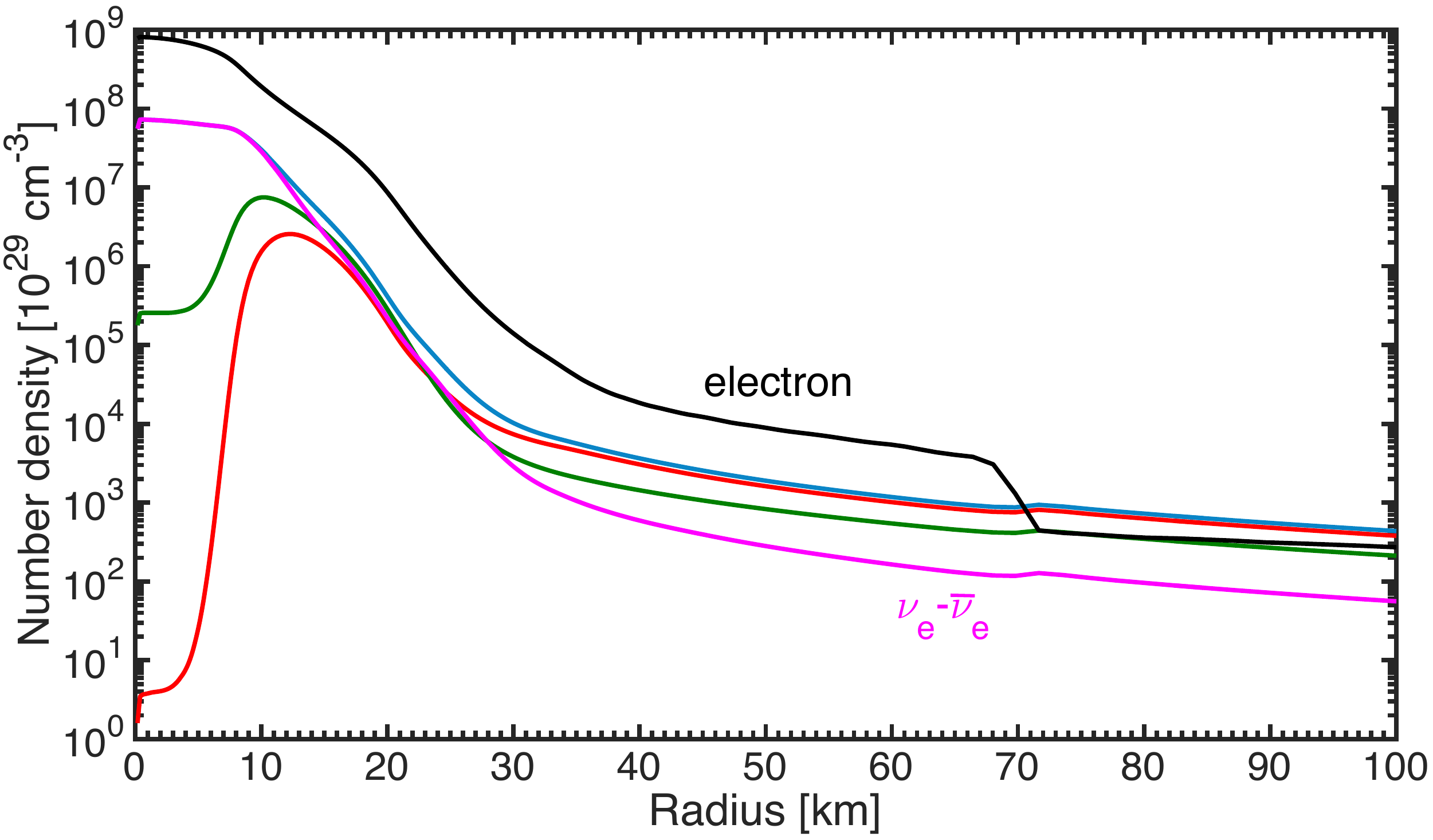}\kern1em\includegraphics[width=0.95\columnwidth]{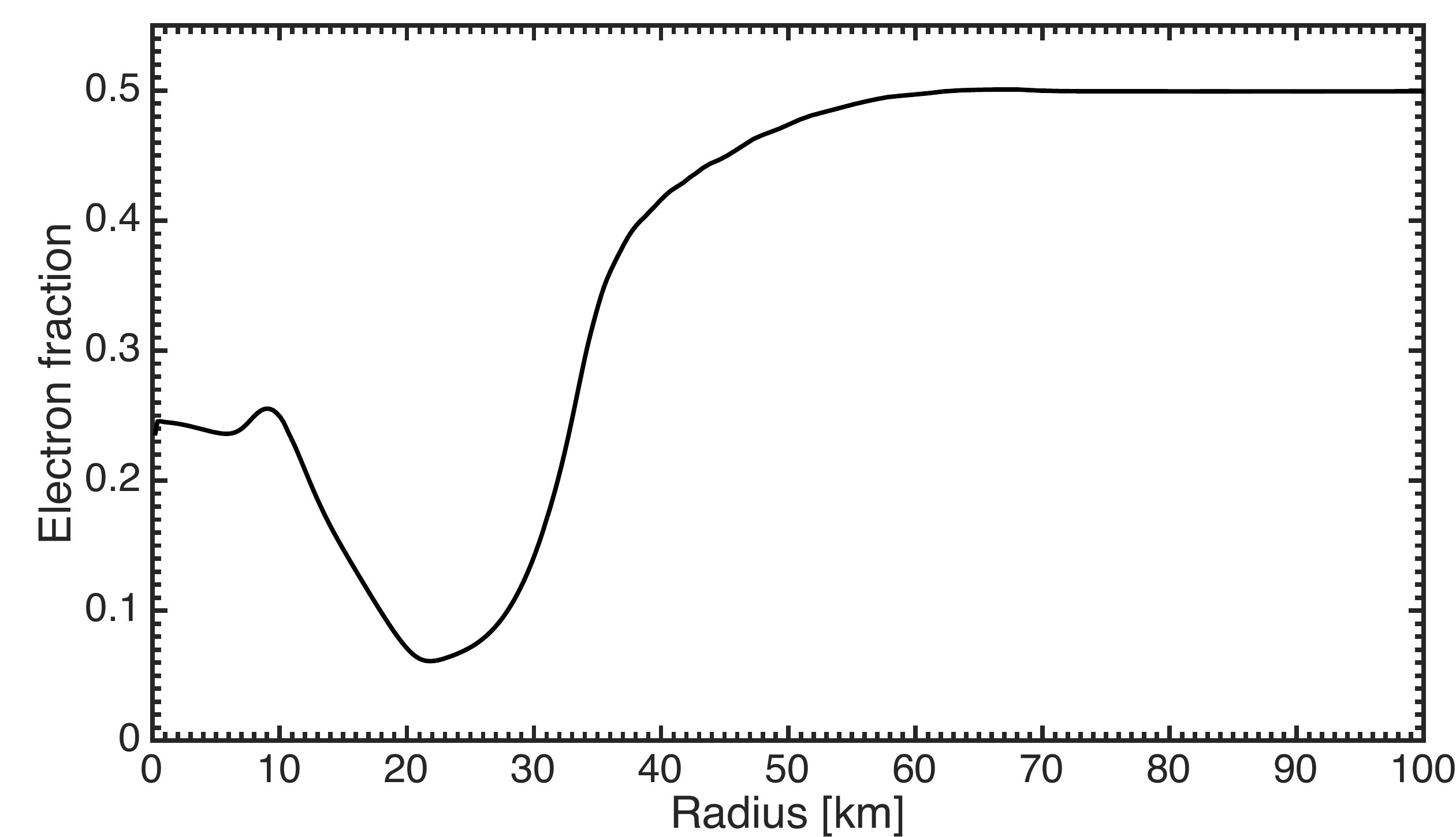}
\caption{Radial variation of global properties of our benchmark case,
the 280~ms snapshot of the $15\ M_\odot$ model.
The blue, red and green lines are for $\nu_e$, $\bar{\nu}_e$ and $\nu_x$ as
indicated. {\em Left panels:} Neutrino luminosity, number luminosity (or
total number flux), and local number density. {\em Right panels:} Matter
velocity, baryon-mass density, and electron fraction (number of electrons
per baryon).
All quantities are given in the comoving fluid frame, so
the discontinuity of the fluid velocity at the shock-wave radius at about
$75$~km imprints itself on other quantities by appropriate red-shift factors.}
 \label{fig:radial_evolution15}
\end{center}
\end{figure*}

\section{Radial variation of neutrino radiation field}\label{sec:radial}

As a first overview, we show the radial variation of several global
properties of the neutrino radiation field in our benchmark model, the 280~ms
snapshot of the $15\,M_\odot$ simulation. The flux density of energy (${\rm
MeV}~{\rm cm}^{-2}~{\rm s}^{-1}$) for a given species $\nu_\alpha$ at a given
radius $r_i$ is
\begin{equation}
\label{eq:number}
{F}(r_i)= 2 \pi \sum_{k=1}^{N_E} \sum_{j=1}^{N_{\mu}} \Delta\mu_{ij} {I}_{ijk} \mu_{ij}\,,
\end{equation}
where $\Delta\mu_{ij}$ is the width of the $\mu=\cos\theta$ bin centered
on $\mu_{ij}$ and the factor $2\pi$ arises from the azimuth integration
\smash{$\int d\varphi$}. The corresponding number flux density
\smash{$\tilde{F}(r_i)$} is the same expression with
$\tilde{I}_{ijk}={I}_{ijk}/E_k$ under the sum. The overall neutrino
luminosity of species $\nu_\alpha$ \smash{(${\rm MeV}~{\rm s}^{-1}$)} at
radial position $r_i$ is \smash{$L(r_i)=4 \pi r_i^2 F(r_i)$} and the
analogous expression for the number luminosity or number flux \smash{(${\rm
s}^{-1}$)} is \smash{$\tilde{L}(r_i)=4 \pi r_i^2 \tilde{F}(r_i)$}. The local
neutrino number density (${\rm cm}^{-3}$) is
\begin{equation}
\label{eq:density}
n(r_i)= 2 \pi c^{-1} \sum_{k=1}^{N_E} \sum_{j=1}^{N_{\mu}} \Delta\mu_{ij} \tilde{I}_{ijk}\,.
\end{equation}

\begin{figure}
\begin{center}
 \includegraphics[width=0.95\columnwidth]{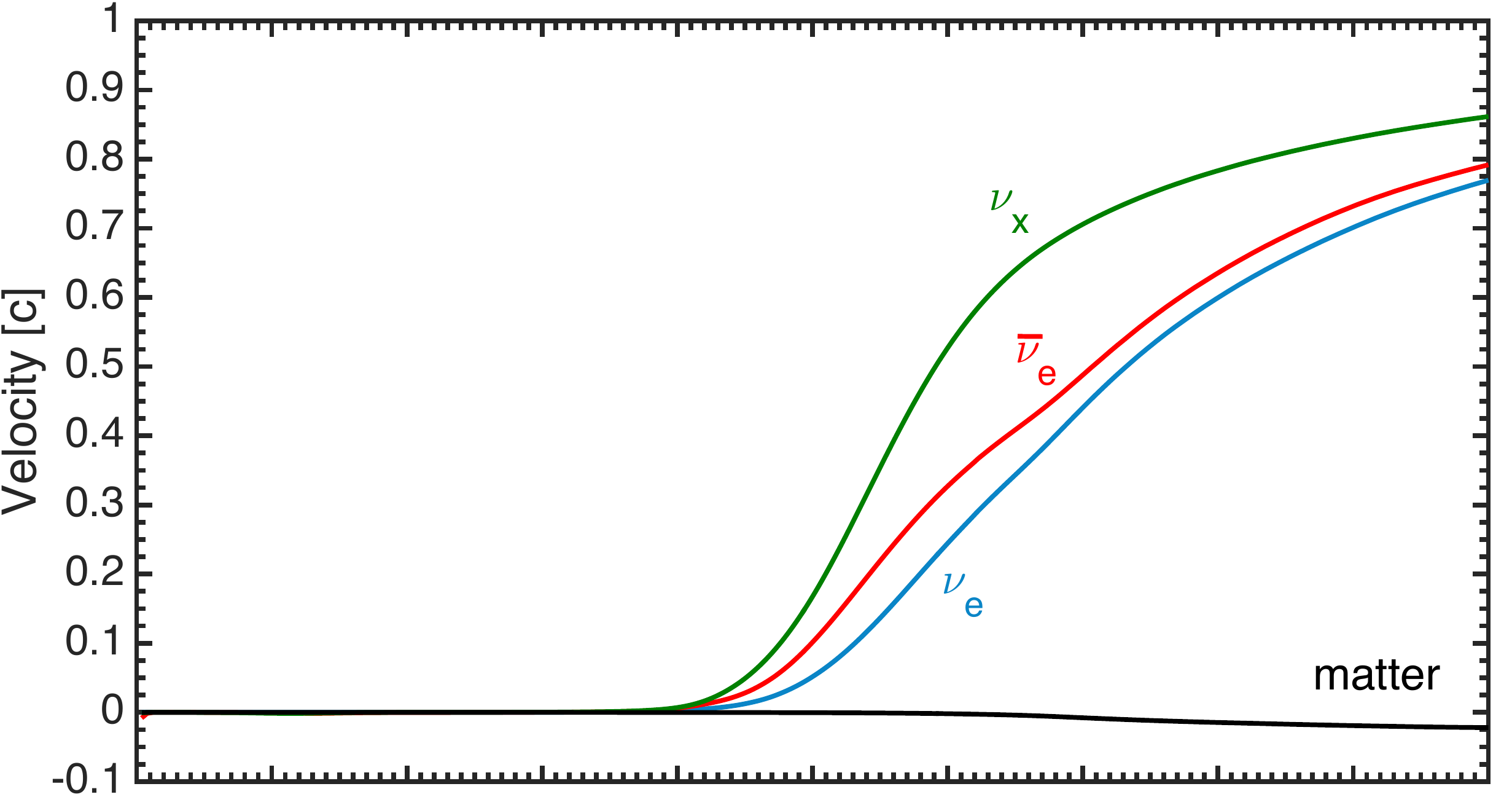}\\[1ex]
 \includegraphics[width=0.95\columnwidth]{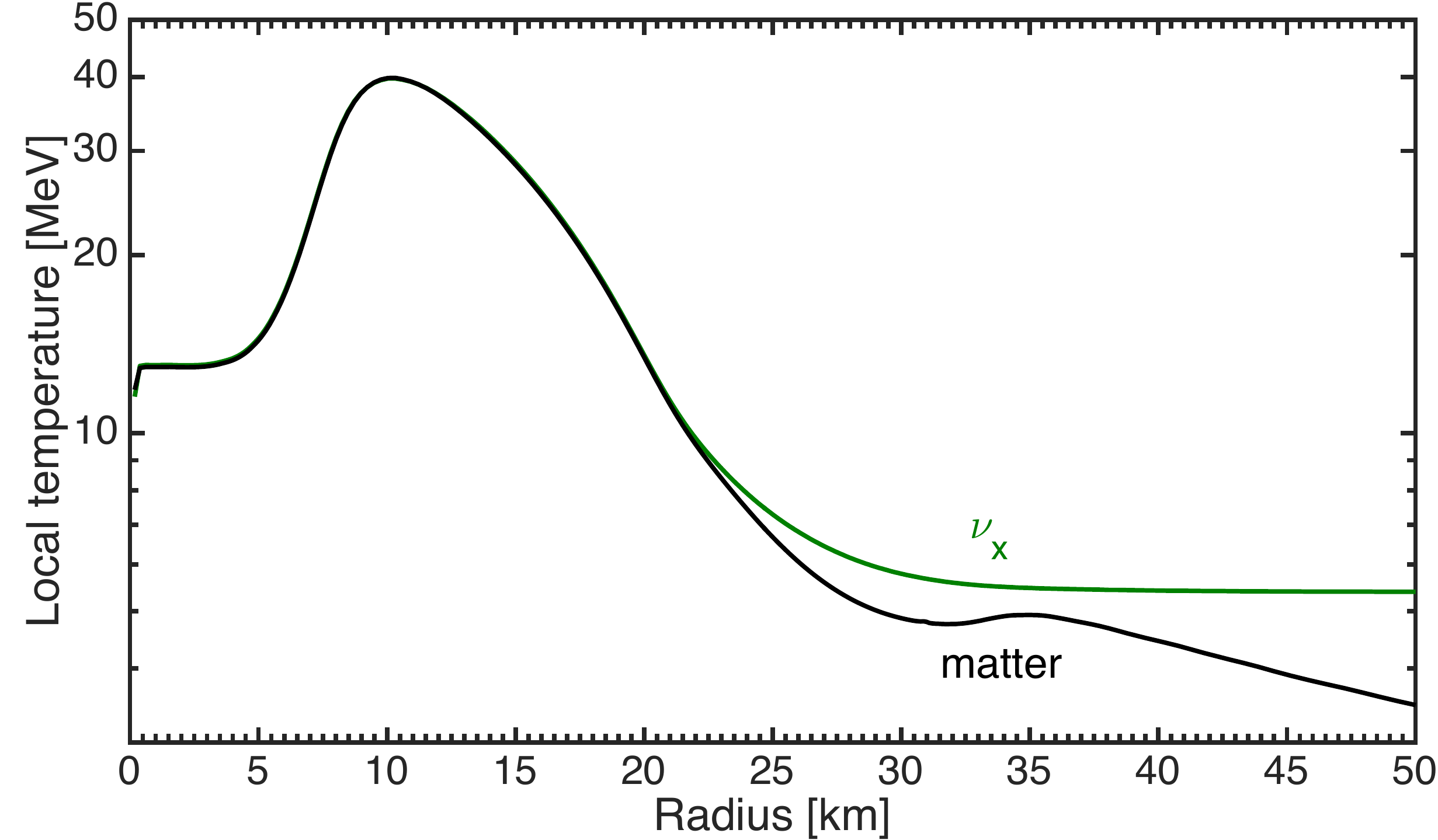}
  \caption{Radial variation of further properties of our benchmark model,
  supplementing Fig.~\ref{fig:radial_evolution15} on a zoomed
  radial scale. {\em Top panel:} Effective neutrino velocity as defined in
  Eq.~(\ref{eq:nuvelocity}), normalized by the speed of light (``flux factor''),
  and the fluid velocity of the matter as in the
  upper-right panel of Fig.~\ref{fig:radial_evolution15}.
  {\em Bottom panel:} Local effective temperature of $\nu_x$ as defined
  in Eq.~(\ref{eq:temperature}) and of the matter background.}
  \label{fig:nuvelocity}
\end{center}
\end{figure}

Figure~\ref{fig:radial_evolution15} shows the radial variation of these and
other characteristics, where quantities for $\nu_e$ are shown in blue, for
$\bar{\nu}_e$ in red, and for $\nu_x$ in green.
The upper left panel shows the flavor-dependent neutrino luminosities. Within
the proto-neutron star, they show fast variations and are also negative in
some region, indicating inward-flowing energy in agreement with the
temperature profile shown in Fig.~\ref{fig:nuvelocity} which has a maximum at
around 10~km. The radial variation depends strongly on flavor in agreement
with the usual picture of flavor-dependent neutrino production and transport.
The sudden increase at the shock-wave radius of around 75~km is an artifact
of expressing the luminosities in the comoving fluid frame. The large infall
velocity of around $-0.2\,c$ outside of the shock wave (upper right panel)
introduces a significant blue shift of both the neutrino energies and their
rate-of-flow. From the perspective of a distant observer in the laboratory
frame, the luminosities
remain continuous through the shock-wave region and are essentially constant
beyond the decoupling region except for smaller changes (of the order of a few
percent) due to energy deposition in the gain layer
below the shock wave at around 75~km.

The $\nu_x$ flavor decouples at a smaller radius than $\nu_e$ and $\bar\nu_e$
and it reaches its shoulder in the luminosity profile at around 20~km. Also
in this case, a further luminosity decline of a few percent (in the
distant-observer frame) occurs from the energy transfer by inelastic
neutrino-nucleon scattering of neutrinos on the cooler stellar plasma between
the energy sphere and the transport
sphere~\citep{Raffelt:2001kv,Keil:2002in}.

The neutrino number flux (or number luminosity) $\tilde L$ is shown in the
left middle panel of Fig.~\ref{fig:radial_evolution15}. Qualitatively, its
radial variation follows the luminosities. However, while the $\nu_e$ and
$\bar\nu_e$ luminosities are nearly equal, the number fluxes differ by the
$\nu_e$ deleptonization flux, compensated by lower average $\nu_e$ energies.

The local electron and neutrino number densities and the electron-neutrino lepton
 ($\nu_e-\bar{\nu}_e$) number density  are shown in the lower left panel. From this plot 
 one concludes that within the shock-wave radius all matter effects are certainly dominated 
 by electrons, which, however,
does not necessarily preclude neutrino self-induced flavor conversion.

The electron fraction (bottom right panel) is locally defined as the ratio of
the net electron number density to the proton plus neutron number density.
Its profile reflects the deleptonization evolution of the infalling matter
of the stellar core with the well-known trough around the neutrinospheric
region and the continuous rise from the corresponding minimal value to
the value of 0.5 (reached at 55~km) of the surrounding layers of the progenitor 
star.

A quantity which nicely illustrates neutrino decoupling is the effective
outward velocity of the neutrino fluid
\begin{equation}
\label{eq:nuvelocity}
v_{\rm eff}(r_i)=\frac{\tilde{F}(r_i)}{n(r_i)}\,,
\end{equation}
which is the number flux density divided by the number density. The ratio
of $v_{\rm eff}$ to the speed of light, $f = v_{\rm eff}/c$ (displayed in
the upper panel of Fig.~\ref{fig:nuvelocity}), is often considered as the
``flux factor'' in neutrino-transport discussions.
Figure~\ref{fig:nuvelocity} (upper panel) shows this quantity as a function
of $r$ for the three species and also the fluid velocity of matter, which is
the same as in the upper right panel of Fig.~\ref{fig:radial_evolution15}.
Deep in the core, where the neutrino distribution is basically isotropic,
$v_{\rm eff}$ is zero, whereas at large radii, where all neutrinos stream
essentially in the radial direction with the speed of light, it
approaches~$c$. In agreement with Fig.~\ref{fig:radial_evolution15}, the
decoupling happens in a flavor-dependent way and neutrinos start to decouple
from the matter fluid at around 22~km. However, while $\nu_x$ decouple almost
instantaneously, $\nu_e$ and $\bar\nu_e$ decouple over a much broader radial
interval, in agreement with the upper left panel of
Fig.~\ref{fig:radial_evolution15}.

Another quantity that illustrates neutrino decoupling is the average neutrino
energy at a given radius, where we mean the average $\langle E\rangle_0$
defined in Eq.~(\ref{eq:firstmom0}), which is the local neutrino energy
density divided by the local number density without weighting with the angular
projection factors $\mu$. For $\nu_x$ with vanishing chemical potential we define
an effective temperature as
\begin{equation}
\label{eq:temperature}
T(r_i)=\frac{180\,\zeta_3}{7\pi^4}\,\langle E\rangle_0\simeq
\frac{\sum_{k=1}^{N_E} \sum_{j=1}^{N_{\mu}} \Delta\mu_{ij} I_{ijk}}{3.15
\sum_{k=1}^{N_E} \sum_{j=1}^{N_{\mu}} \Delta\mu_{ij} \tilde{I}_{ijk}}\,.
\end{equation}
This quantity and the matter temperature are plotted in the
bottom panel of Fig.~\ref{fig:nuvelocity}. At $r\simeq 22$~km, the $\nu_x$
start to decouple from the stellar material and their temperature differs
relative to the one of the surrounding matter. The effective $T$ of the
final $\nu_x$ flux is considerably lower than it is at the ``energy
sphere'' where $\nu_x$ begins to decouple; see \cite{Keil:2002in} for more
details.
\vspace{1cm}

\section{Angular variation of neutrino radiation field}\label{sec:angles}

In this section, we finally explore the neutrino angle distributions for all
flavors, and in particular, how they vary with radius and evolve in time. We
also observe that the local neutrino spectrum in a given direction of
propagation is well described by a Gamma distribution.

\subsection{Radial variation and temporal evolution of the neutrino angle distributions}

In order to grasp the general trend of how the neutrino angular distributions
vary in space and time we introduce at a given radius the local number
intensity (see also Eq.~\ref{eq:locdens}),
\begin{equation}
\tilde{I}_\mu = 2\pi\sum_{k=1}^{N_E} \tilde{I}_{ijk}\,,
\end{equation}
which is the local number density of streaming neutrinos, 
integrated over energy and azimuth angle, but
differential with regard to $\mu=\cos\theta$, and has units of ${\rm
cm}^{-2}~{\rm s}^{-1}$. 
More specifically, in what follows, we will show the corresponding
number density (not the number intensity), i.e., $\tilde{I}_{\mu}/c$,
which has units \smash{${\rm cm}^{-3}$}.

\begin{figure}
\begin{center}
  \includegraphics[width=\columnwidth]{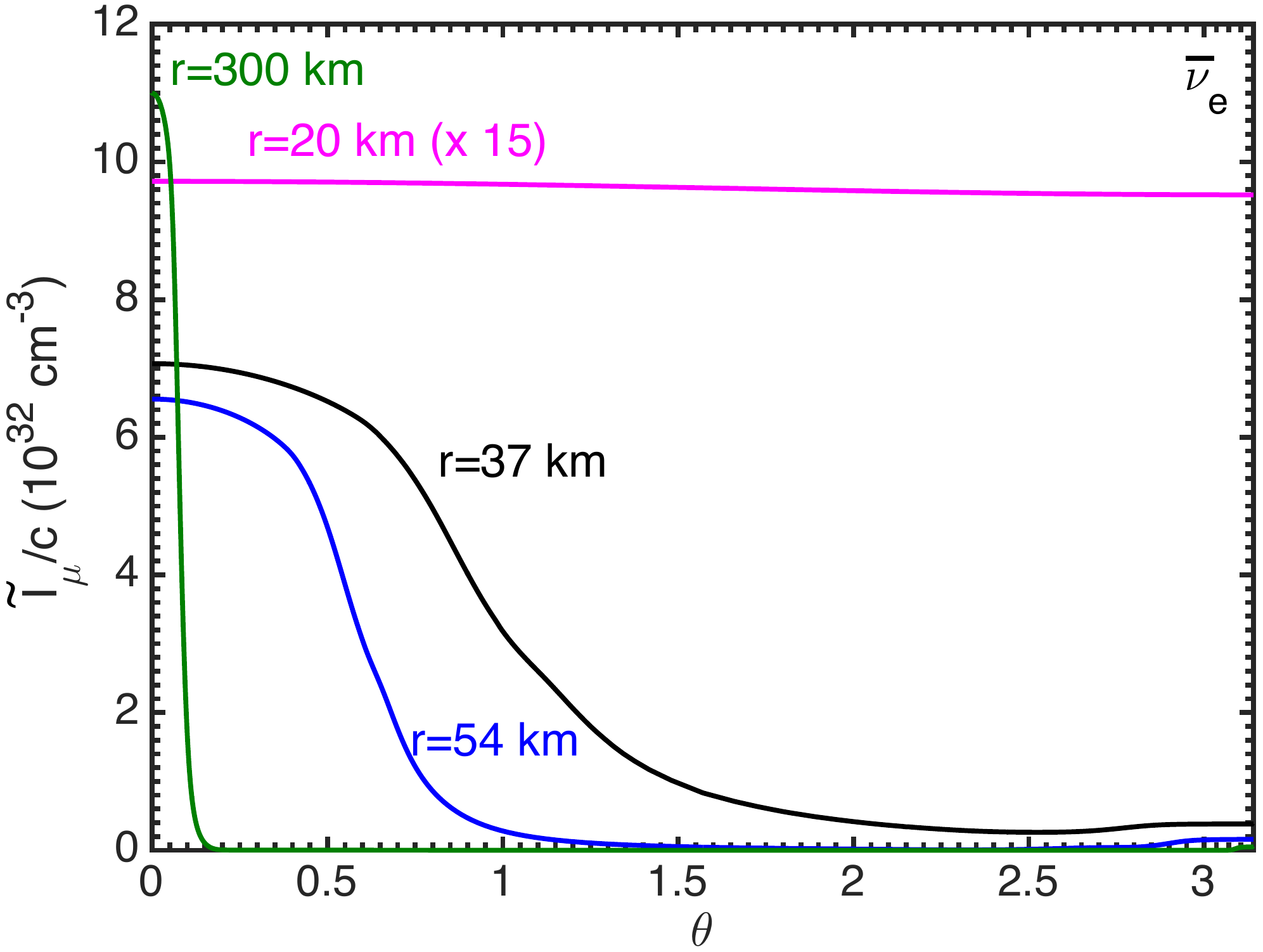}\\
 \hspace{-2.6mm}\includegraphics[width=1.03\columnwidth]{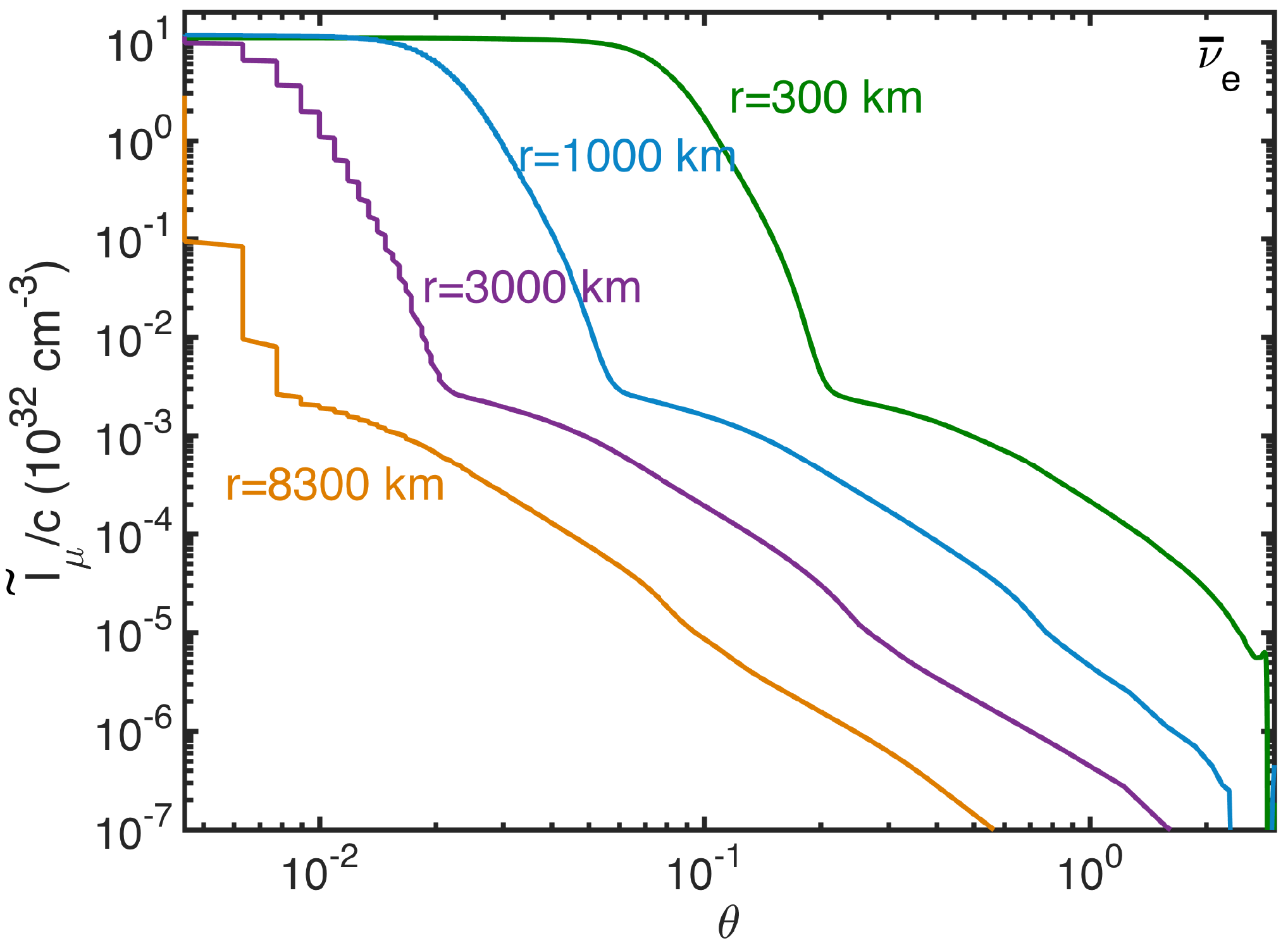}
  \caption{Number intensities for $\bar\nu_e$ as a function of zenith angle
  for the indicated radial distances of the 280~ms snapshot of our
  $15\,M_\odot$ model. {\em Top:} Linear plot for small radii.
  The angular distributions become more forward peaked for larger radii, but
  even at 37~km (black curve) they still stream in all directions.
  {\em Bottom:} Log-log-plot for larger radii, revealing the forward-peaked
  ``neutrino beam'' and a broad ``halo'' arising from residual scattering
  of the beam.}
   \label{fig:angular_distributions}
\end{center}
\end{figure}

Figure~\ref{fig:angular_distributions} shows  $\tilde{I}_{\mu}/c$ for $\bar{\nu}_e$ as a 
function of $\theta$ for the indicated radial distances. The upper panel shows the distributions at small radii. As expected, inside
the decoupling region the distribution is almost isotropic (see the magenta
curve at $r=20$~km). At larger radii, the angle distributions become more and
more forward peaked, corresponding to $\theta = 0$. However, even at
$r=37$~km (black curve), a non-negligible backward contribution remains
clearly visible in this linear plot, confirming that neutrinos stream fully
in the forward direction only for $r > 40$~km, as speculated from
Fig.~\ref{fig:radial_evolution15}. For $r=54$~km (blue curve), the
distribution already becomes similar to a beam with a broad opening angle
that becomes smaller with distance from the source, corresponding to the
angular size of the SN core as seen from the given distance.

A logarithmic plot (bottom panel of
  Fig.~\ref{fig:angular_distributions}), however, reveals that this
  picture is not complete. As neutrinos stream outward, they suffer
  residual collisions with the matter layers. As a consequence, the
  ``neutrino beam'' emerging from the source is accompanied by a
  ``halo'' which extends to all directions. The neutrino halo
  populating the large $\theta$ phase space is visible in the bottom
  panel of Fig.~\ref{fig:angular_distributions}.  One can see as the
  number intensity of the neutrino halo is small with respect to the
  neutrino beam propagating forward ($\theta \simeq 0$), but its broad
  angular distribution allows it to dominate the neutrino-neutrino
  interaction energy as discussed by \cite{Sarikas:2012vb}.  


\begin{figure}
\begin{center}
  \includegraphics[width=\columnwidth]{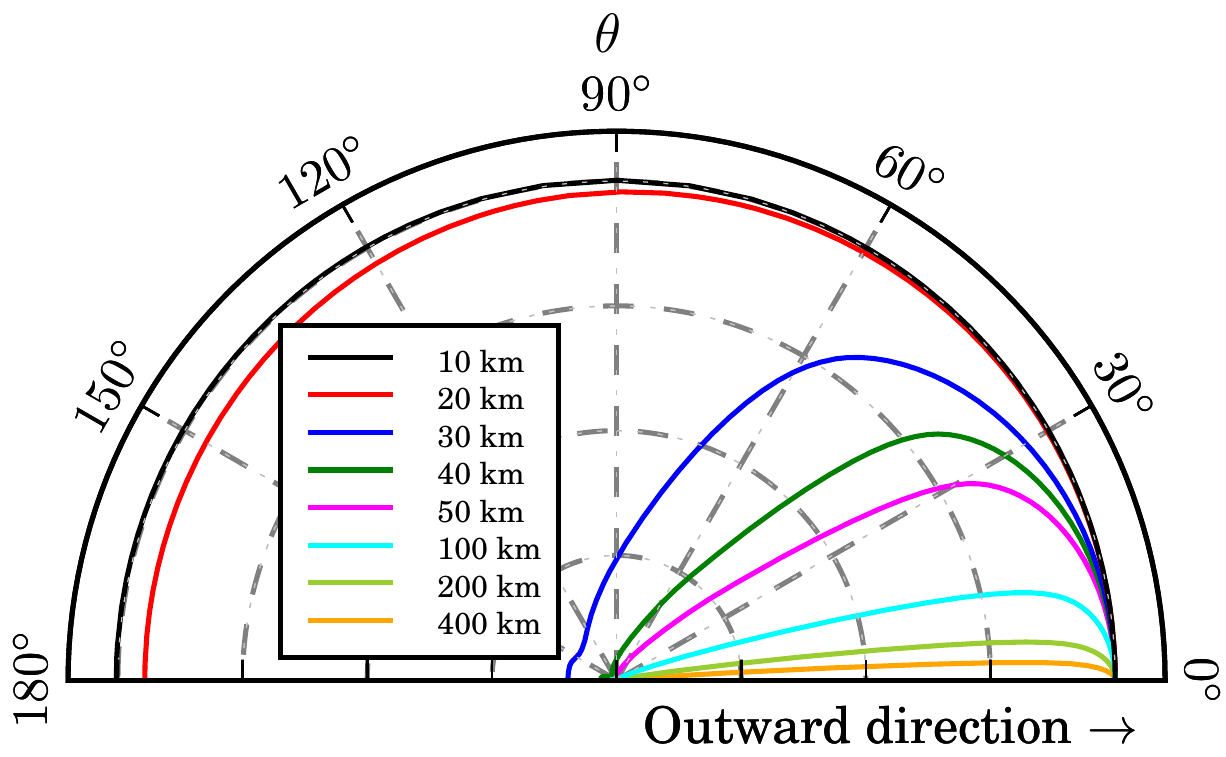}
  \caption{Radial variation of the 
  normalized spectral intensity $I_{E,\Omega}$ of $\bar{\nu}_e$ for $E=10$~MeV and
  the 280~ms snapshot of our
  $15\,M_\odot$ model as a function of the zenith direction angle. The angle
  $\theta =0$ refers to the local radial direction.
 }
   \label{fig:polarmap}
\end{center}
\end{figure}

Figure~\ref{fig:polarmap} is a polar map of the normalized $\bar{\nu}_e$
monochromatic intensity $I_{E,\Omega}$ introduced in Eq.~(\ref{eq:monoint})
for $E_k=10$~MeV and at different radii. For larger radii, the distributions
become more forward peaked, and outside of the neutrino-decoupling region
the intensity is essentially conserved along radiation paths that point outward
from the radiating neutrinosphere.
A similar behavior was also found by
\cite{Thompson:2002mw}, see their Figs.~4 and 10--12.

\begin{figure}
\begin{center}
  \includegraphics[width=\columnwidth]{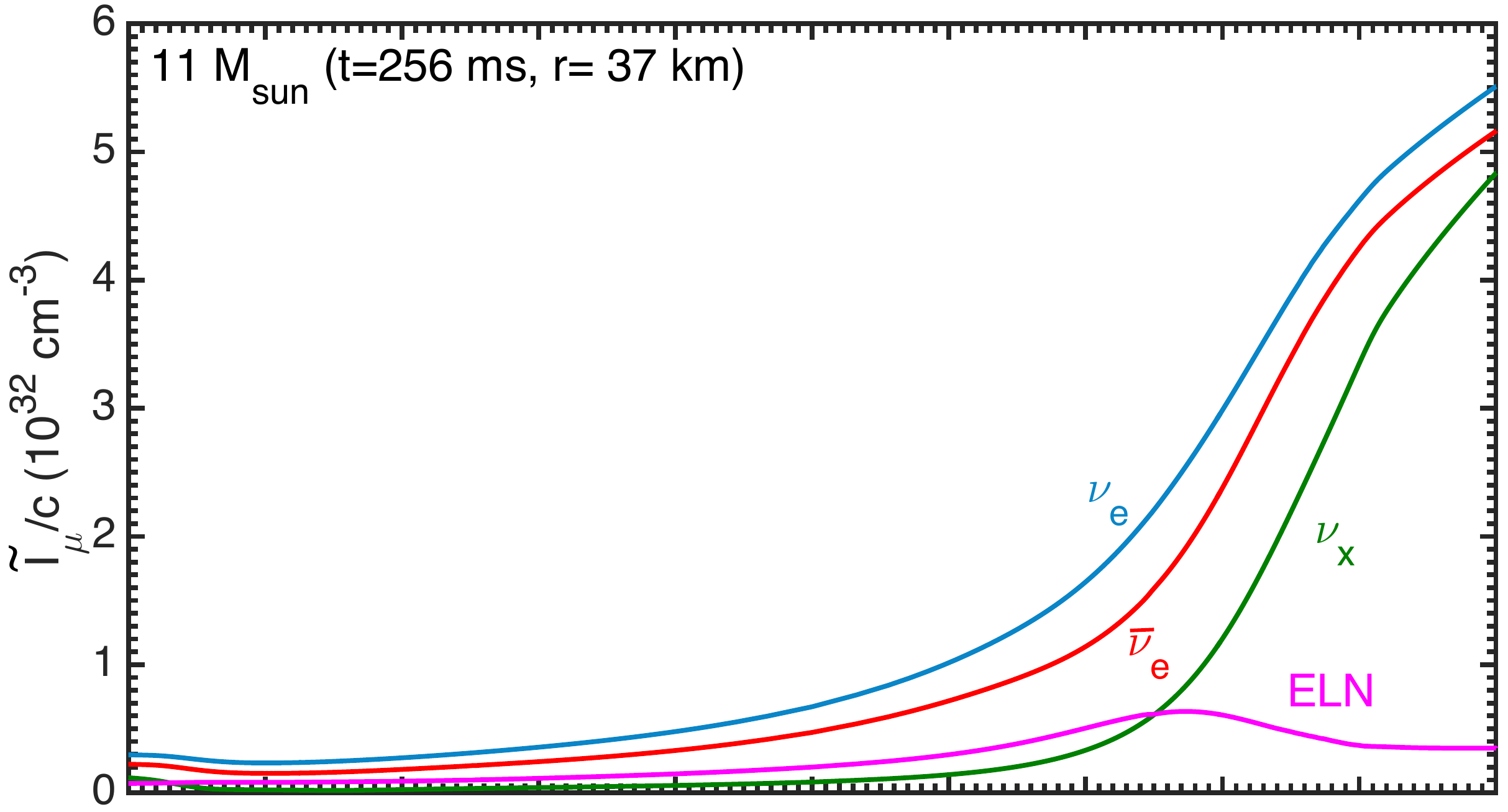}  \\
  \includegraphics[width=\columnwidth]{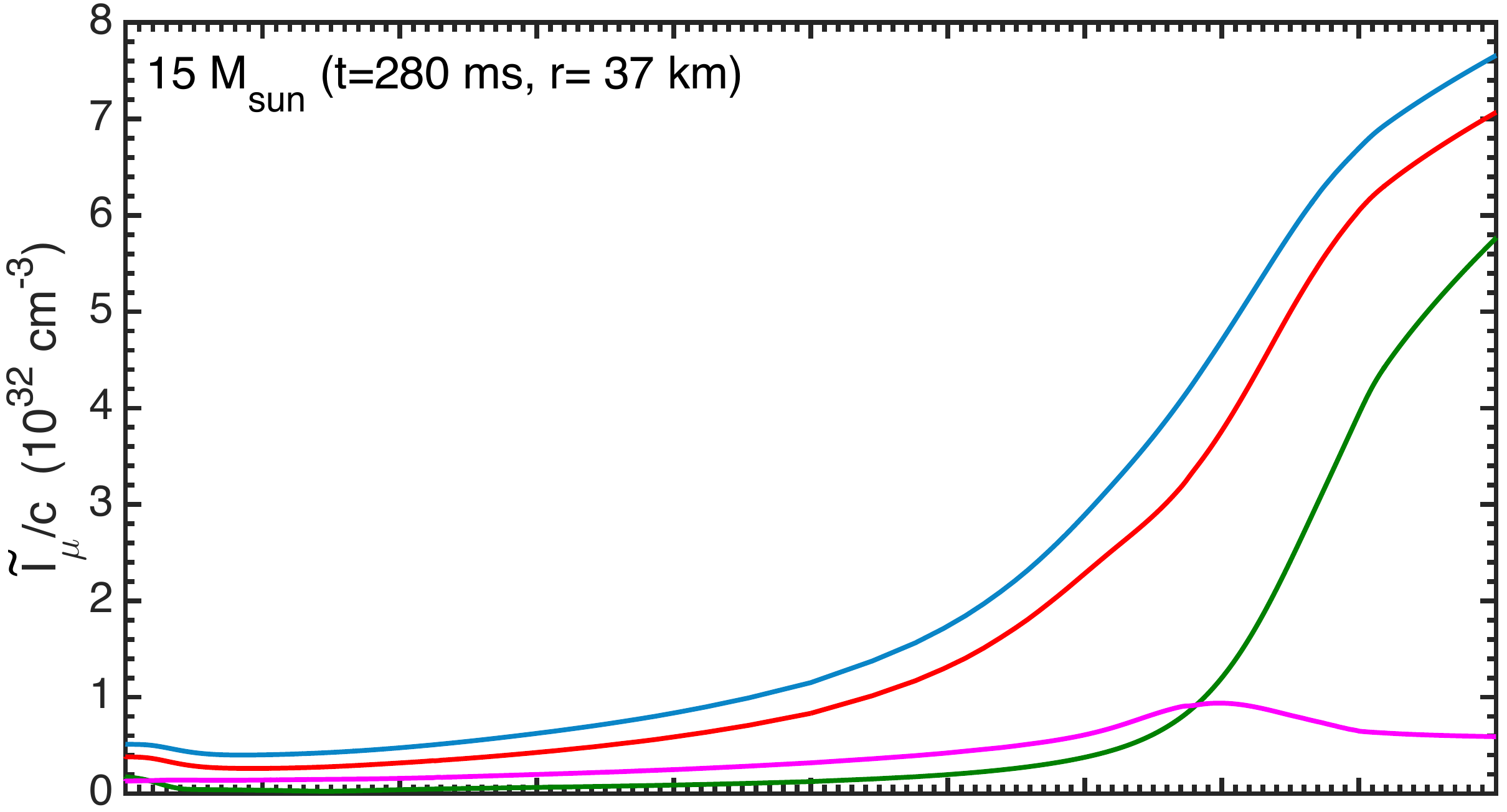} \\
\hspace{-2.3mm} \includegraphics[width=1.02\columnwidth]{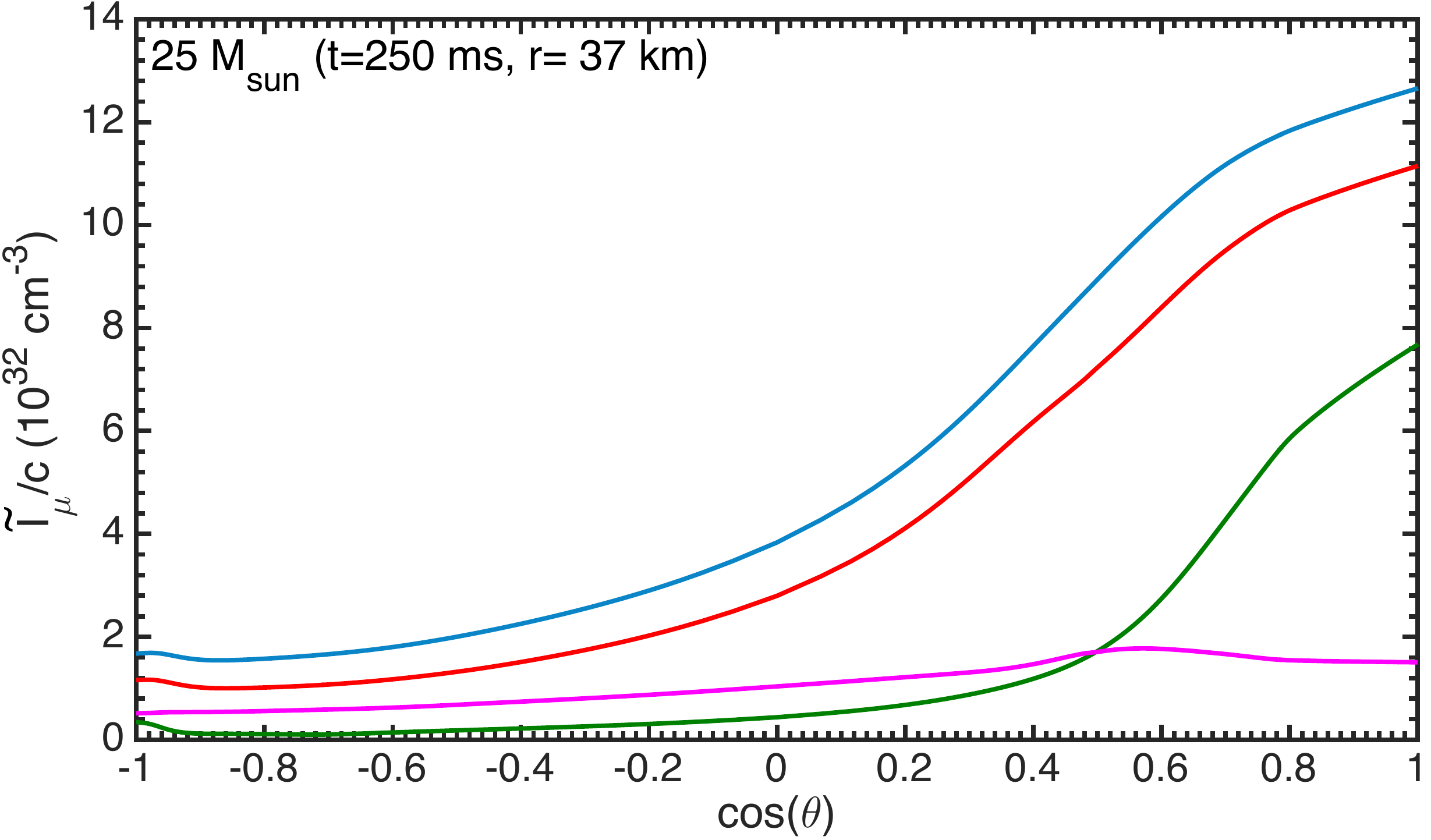}
  \caption{Neutrino number intensity as a function of $\cos\theta$ for the
  species $\nu_e$ in blue, $\bar{\nu}_e$ in red, $\nu_x$ in green, and $\nu_e-\bar{\nu}_e$ 
  in magenta as indicated.
  From top to bottom for snapshots at comparable times at 37~km for our three
  progenitor models as indicated in the panels. As expected, the
  height and width of the distributions decreases in the sequence
  $\nu_e$, $\bar\nu_e$ and $\nu_x$.}
  \label{fig:angular_evolution}
\end{center}
\end{figure}

Figure~\ref{fig:angular_evolution} shows the flavor dependent angle
distributions for our three progenitors. The width of the distributions
decreases in the sequence $\nu_e$, $\bar\nu_e$ and $\nu_x$
\citep{Ott:2008jb,Sarikas:2011am} as expected from the decreasing interaction
rates in the medium. Qualitatively, this behavior is the same for all
progenitors. At some radii, the angular distributions of $\nu_e$ and $\nu_x$ or
the ones of $\bar{\nu}_e$ and $\bar{\nu}_x$ may cross as assumed in
\cite{Mirizzi:2011tu}.

\subsection{Neutrino electron lepton number}\label{sec:oscillations}

A new issue in the context of self-induced neutrino flavor conversion is the
question of the angular distribution of the electron lepton number (ELN)
carried by neutrinos and in particular the question of possible crossings of
the $\bar{\nu}_e$ with the $\nu_e$ angle distributions
\citep{Sawyer:2015dsa,Izaguirre:2016gsx}. Such situations could trigger fast
flavor conversion, i.e., self-induced flavor conversion where the instability
scale is not set by the neutrino mass differences, but rather by the
neutrino-neutrino interaction energy, a much larger scale for typical SN
conditions.

\begin{figure}
\begin{center}
  \includegraphics[width=\columnwidth]{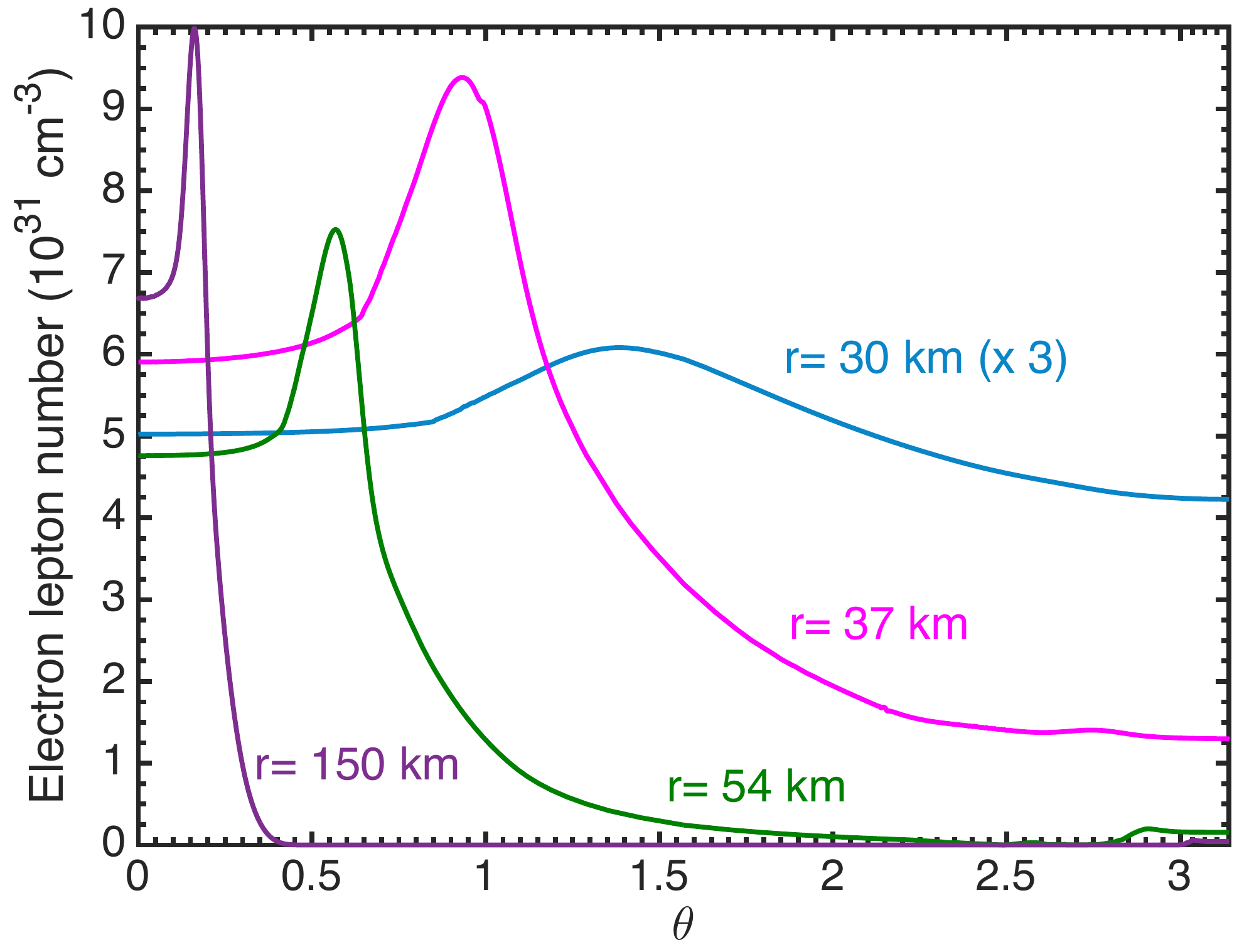}
  \caption{Angular distribution of the neutrino electron lepton number for the
  280~ms snapshot of the $15\,M_\odot$ progenitor extracted at different
  radii as indicated.
  }
   \label{fig:ELN}
\end{center}
\end{figure}

The ELN carried by neutrinos, defined as
$(\tilde{I}_{\mu,\nu_e}-\tilde{I}_{\mu,\bar{\nu}_e})/c$, is shown in magenta in Fig.~\ref{fig:angular_evolution}
for the selected snapshots of our three progenitors. It is comparable in intensity for all three progenitors, but
the dip in the forward direction is less pronounced for the $25\,M_\odot$ SN model for the selected post-bounce time
and radius.
Figure~\ref{fig:ELN} shows the ELN as a function of $\theta$ for different radii for our 
benchmark model. It
develops a  dip in the forward direction as the radius increases. We also
provide an animation of the ELN variation with radius for the $15\,M_\odot$
progenitor \href{http://wwwmpa.mpa-garching.mpg.de/ccsnarchive/data/Tamborra2017/}
{here}. In all of our models we have found
the ELN to be always positive. In particular, the universal dip in the
forward direction never turns negative, so there is no crossing between the
$\nu_e$ and $\bar\nu_e$ angular distributions. On the contrary, crossings in the ELN
distribution naturally appear in compact merger
remnants because of the different emission geometry with respect to SNe and the 
$\bar{\nu}_e$ flux being larger than the $\nu_e$ one~\citep{Wu:2017qpc}.

This behavior could change in 3D models, for example in the
presence of LESA 
\citep[Lepton-Emission Self-sustained Asymmetry;][]{Tamborra:2014aua}. 
 LESA manifests itself in a 
  pronounced large-scale dipolar pattern in the ELN emission.
  This is a consequence of large-scale convection modes 
  inside the newly formed neutron star with a strong dipolar
  flow component, which grows during the contraction phase~\citep{Janka:2016fox}
  and has significant feedback also on the accretion
  flow around the neutron star \citep{Tamborra:2014aua}. 
LESA naturally implies a change of sign in
  $(\tilde{I}_{\mu,\nu_e}-\tilde{I}_{\mu,\bar{\nu}_e})$.  It is
  therefore conceivable that, especially in the regions where the ELN
  changes its sign, crossings in the ELN angular distributions may
  occur.

 Existing 3D simulations employ, however, the
  ray-by-ray transport approximation~\citep[]{Melson:2015tia,Melson:2015spa,2015ApJ...807L..31L,Takiwaki:2013cqa}
  or multi-dimensional two-moment treatments with algebraic
  closure relations such as the ``M1 methods''~\cite[]{Roberts:2016lzn}. 
  None of these can
  provide reliable angle distributions. M1 solvers use the
  neutrino energy and momentum equations to evolve
  the {\em angle-integrated} moments of the neutrino intensity
  (i.e., the energy-dependent energy and flux densities).
  The ray-by-ray approximation, even if based on a two-moment solver
  with Boltzmann closure \citep{RamppJanka2002,Burasetal2006},
  makes use of the assumption that the neutrino intensity is
  locally axially symmetric around the radial direction. It
  thus ignores non-radial flux components and off-diagonal
  elements in the pressure tensor. 
  For certain questions 
  (e.g.\ for rough estimates of observable emission asymmetries)
  one can work around the missing phase-space information by
  using low-order multipole approximations of the intensity
  based on energy-density and flux-density information~\citep[]{Muller:2011yi,Tamborra:2014hga}. Detailed, accurate phase-space information,
  however, requires the solution of the time-dependent 
  Boltzmann transport equation for the neutrino intensity
  in dependence on all three spatial and all three momentum-space
  variables. 
  
  Existing supercomputers can hardly tackle this 
  high-dimensional problem in two spatial dimensions
  despite severe restrictions
  with respect to resolution and sophistication
  of the employed weak-interaction physics~\citep{Ott:2008jb,Brandt:2010xa,Nagakura:2017mnp}.
  Time-dependent numerical solutions of the Boltzmann transport 
  equation in 3D will require exascale computing capability,
  but will still be very challenging when high resolution in 
  momentum space is required. The results based on a tangent-ray
  integration of the energy and angle-dependent Boltzmann 
  equation in 1D presented in this work are therefore likely
  to set the benchmark for neutrino-oscillation studies for the
  coming years.

\subsection{Spectral fit}
\begin{figure}
\begin{center}
 \includegraphics[width=.985\columnwidth]{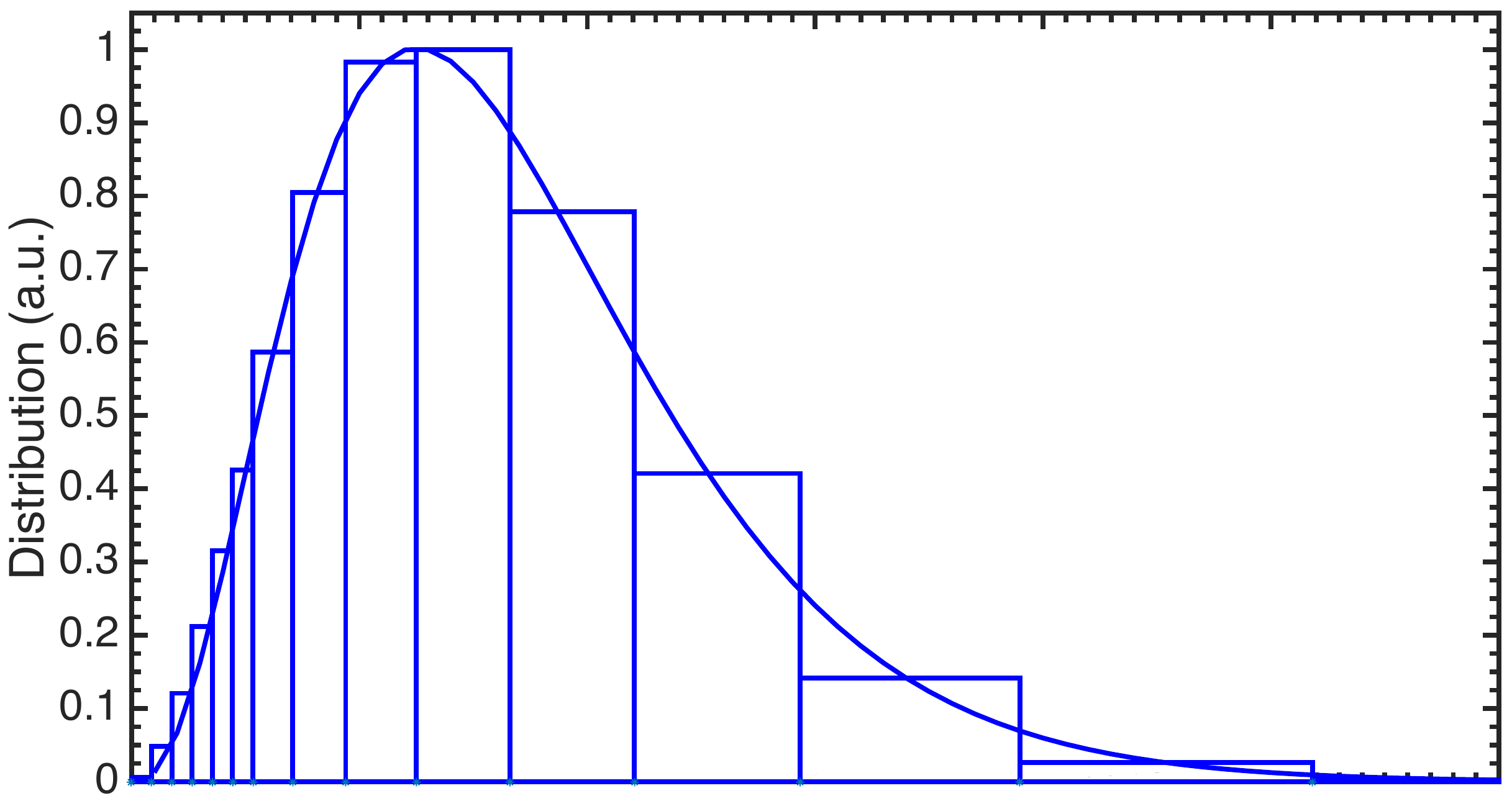}\\
\includegraphics[width=1.\columnwidth]{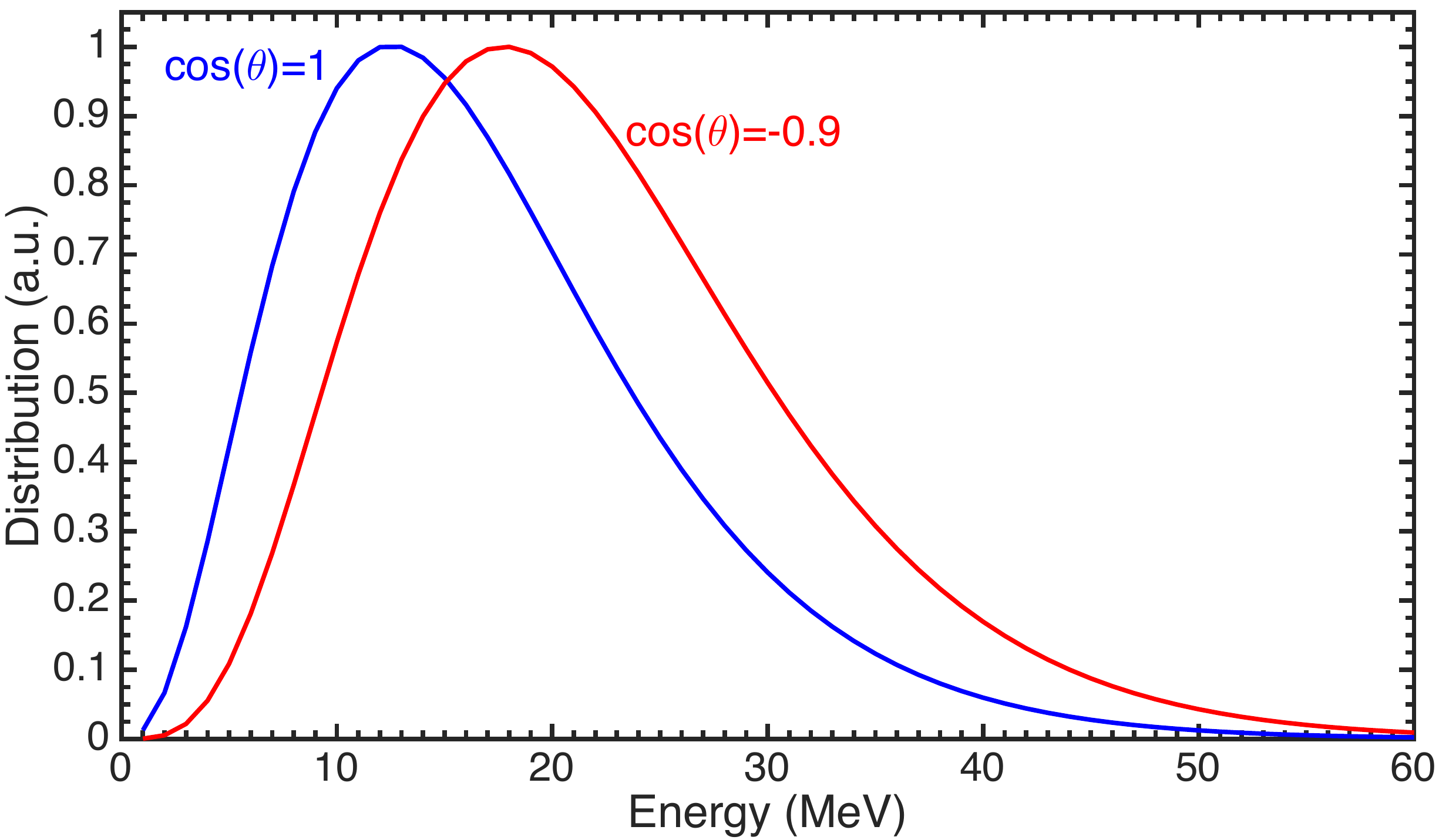}
  \caption{Representative example for the $\bar{\nu}_e$ spectrum (energy distribution) at
  $r=30$~km for the 280~ms snapshot of the $15\,M_\odot$ model, where blue
  is for $\cos(\theta)=1$ and red for $\cos(\theta)=-0.9$.
  {\em Top panel:} Normalized histogram of the numerical spectrum and
  fit by a Gamma distribution (smooth line).
  {\em Bottom panel:} Same fit (blue line) as well as a similar fit
  for a nearly backward direction, both of them normalized to the same value
  at maximum. The backward flux has larger energies caused by the energy
  dependence of the neutrino cross sections.}
   \label{fig:alpha_fit}
\end{center}
\end{figure}

Flavor oscillation effects generally depend both on neutrino energy and
angle. Therefore, in our data files we also provide information about the
energy distribution. It has been observed previously that SN neutrino energy
spectra often can be well approximated by a Gamma distribution
~\citep{Keil:2002in,Tamborra:2012ac}, sometimes also referred to as ``$\alpha$
fit.'' In our context it means that we will express the spectral intensities
in the form 
\begin{equation}
\label{eq:alphafit}
I_{E,\theta}\propto \left(\frac{E}{E_\theta}\right)^{\alpha_\theta}
e^{-(\alpha_\theta+1) E/E_\theta}\,,
\end{equation}
where the energy $E_\theta=\langle E\rangle_\theta$ is the average
energy of the neutrinos streaming in direction $\theta$, whereas the shape
parameter $\alpha_\theta$ measures the amount of spectral pinching. It is
related to the first and second energy moments through
\begin{equation}
\frac{\langle E^2\rangle_\theta}{\langle E\rangle_\theta^2} =
\frac{2+\alpha_\theta}{1+\alpha_\theta}\,.
\end{equation}
Besides the energy-integrated intensity $I_\mu$ we \href{http://wwwmpa.mpa-garching.mpg.de/ccsnarchive/data/Tamborra2017/}
{provide}
the first two energy moments (Eqs.~\ref{eq:mom1}, \ref{eq:mom2})
for any $\mu$ and in this way characterize the spectra with
reasonable accuracy.

As a representative example, we show in Fig.~\ref{fig:alpha_fit} the
$\bar\nu_e$ energy spectrum at $r=30$~km for our benchmark SN model. We consider the
outward direction ($\mu=1$ in blue) and a nearly backward direction
($\mu=-0.9$ in red). For the former we show both the numerical distribution
as a histogram and the fit by a Gamma distribution (top panel). We then compare the
smooth fits for both directions (bottom panel of Fig.~\ref{fig:alpha_fit}), illustrating that the 
energy distributions
depend significantly on direction simply because of the energy-dependent
neutrino scattering cross sections. A similar comparison was provided by
\cite{Sarikas:2012vb} at large distances to compare the spectrum of the
neutrinos coming directly from the SN core vs.\ the halo flux which arises
from residual scattering. In our case, we find that the $\bar\nu_e$ streaming
in the forward direction have $E_\theta=17.1$~MeV and
$\alpha_\theta=2.7$ and are therefore cooler and less pinched than the
quasi-backward direction, where $E_\theta=22.1$~MeV and
$\alpha_\theta=4$. More energetic neutrinos are typically more
isotropically distributed in energy than less energetic ones.

\section{Conclusions}\label{sec:conclusions}

Recent developments in the context of neutrino-neutrino interactions in dense
media have highlighted the role of the neutrino angular distributions. For
the first time, we have carried out a careful analysis of the neutrino
angular distributions from one-dimensional hydrodynamical SN simulations with
sophisticated neutrino transport.

To facilitate dedicated studies of flavor conversion, we provide \href{http://wwwmpa.mpa-garching.mpg.de/ccsnarchive/data/Tamborra2017/}
{data} of the
energy and angle distributions as supplementary
material. We specifically explore the angle distributions of neutrinos in
three SN progenitors with masses of 11.2, 15 and $25\,M_\odot$, and scan
three different post-bounce times during the accretion phase for each of
them.

The neutrino radiation field ranges from being completely isotropic inside
the proto-neutron star, where neutrinos frequently scatter, to being
forward peaked at large distances from the proto-neutron star (50~km for the
analyzed models) where neutrinos stream almost freely. However, while
propagating further, neutrinos scatter on nuclei in the stellar envelope,
generating a broad ``halo'' and backward distribution, becoming relevant at larger
radii. The neutrino radiation flowing in the backward direction is in general
characterized by an energy spectrum hotter and more pinched than the forward
streaming component.

The question if fast flavor conversion occurs in the neutrino decoupling
region will require a better theoretical understanding of the behavior of the
neutrino radiation field in the SN environment. Our present study provides
the input information that is required for such studies.

\section*{acknowledgements}

We acknowledge support from the Knud H{\o}jgaard Foundation, the Villum
Foundation (Project No.\ 13164), the Danish National Research Foundation
(DNRF91), the Deutsche Forschungsgemeinschaft under the Excellence Cluster
Universe (Grant No.\ EXC 153), and the European Union under the Innovative
Training Network ``Elusives'' (Grant No.\ H2020-MSCA-ITN-2015/674896) and
under the ERC Advanced Grant No.\ 341157-COCO2CASA. We
thank the Institute for Nuclear Theory at the University of Washington for
hospitality and the DOE for partial support during early stages of
this work.


\appendix

\section{Neutrino Radiation Field} \label{sec:RadiationField}

In the context of neutrino flavor oscillation studies, the local neutrino
radiation field for a given species $\nu_e$, $\bar\nu_e$ or $\nu_x$ is
traditionally described by occupation numbers (classical phase-space
densities) $f({\bf p})$ for every momentum mode ${\bf p}$. These can be
extended to $3{\times}3$ matrices $\varrho({\bf p})$ to capture flavor
coherence on the off-diagonal elements. On the other hand, in
the traditional treatment of neutrino radiative transfer, one uses the
spectral intensity $I_{E,\Omega}$ as the fundamental quantity, i.e., the
energy carried by the given neutrino species per unit area and unit time, 
differential with respect 
to neutrino energy $E$ and solid angle $\Omega$ (see Eq.~\ref{eq:monoint}).
 Moreover, in the particle-physics tradition one uses natural
units with $\hbar=c=1$, although we will keep $\hbar$ and $c$
explicitly here. In this appendix, we provide a brief dictionary between these
languages because simple issues of definition, normalization or units can be
a source of confusion.

Taking neutrinos to be massless, their energy is $E=c|{\bf p}|$. Therefore, the
differential local number density is
\begin{equation}
dn=\frac{f_{\bf p}}{\hbar^3}\,\frac{d^3{\bf p}}{(2\pi)^3}
=\frac{f_{E,\Omega}}{(\hbar c)^3}\,\frac{E^2dEd\Omega}{(2\pi)^3}\,,
\end{equation}
where we show the dependence on ${\bf p}$, $E$, or $\Omega$ as indices.
In the second expression the neutrino momentum is represented by
its energy and direction of motion.
Noting that massless neutrinos stream with the speed of light $c$ in a given
direction $\Omega$, the spectral intensity is
\begin{equation}
I_{E,\Omega}=c E\,\frac{dn}{dEd\Omega}=
\frac{f_{E,\Omega}}{\hbar^3 c^2}\,\frac{E^3}{(2\pi)^3}
\end{equation}
in units of ${\rm cm}^{-2}~{\rm s}^{-1}~{\rm ster}^{-1}$.
Integrating over all energies, the intensity is
$I_\Omega=\int_0^\infty I_{E,\Omega}\,dE$
in units of \smash{${\rm MeV}~{\rm cm}^{-2}~{\rm s}^{-1}~{\rm ster}^{-1}$}.

In the neutrino flavor evolution context, the relevant quantities are number
densities and number fluxes, not energy densities or fluxes. To develop a
systematic notation we here use a tilde on a symbol for the corresponding
number quantity. In particular, we define the spectral number intensity
\begin{equation}
\tilde I_{E,\Omega}=\frac{I_{E,\Omega}}{E}=c\,\frac{dn}{dEd\Omega}=
\frac{f_{E,\Omega}}{\hbar^3 c^2}\,\frac{E^2}{(2\pi)^3}
\end{equation}
in units of \smash{${\rm cm}^{-2}~{\rm s}^{-1}~{\rm MeV}^{-1}~{\rm ster}^{-1}$}.

For producing a weak potential on other neutrinos, a more intuitive quantity
is the local number density, differential with regard to energy and solid angle.
Therefore, without introducing a
special symbol, our real quantity of interest is
$\tilde I_{E,\Omega}/c$
in units of \smash{${\rm cm}^{-3}~{\rm MeV}^{-1}~{\rm ster}^{-1}$}. Of course,
in natural units where $c=1$, both quantities are the same. Still,
there is a conceptual difference between a flux-like quantity
($I_{E,\Omega}$ or $\tilde I_{E,\Omega}$)
and a density, even though for relativistic particles they are equivalent  except for units.

Henceforth we assume that the radiation field is axially symmetric around the
local radial direction. Moreover, we assume that the overall SN model is
spherically symmetric. We use local polar coordinates
$\Omega=(\theta,\varphi)$ with the differential $d\Omega=d\mu\,d\varphi$
with $\mu=\cos\theta$. The azimuthal integration $\int d\varphi\to2\pi$ is
trivial because of the assumed symmetry.

In particular, we are interested in the number intensity, integrated over energy
and azimuth angle, which is
\begin{equation}
\label{eq:locdens}
\frac{\tilde I_\mu}{c}=\int dE\,d\varphi \frac{\tilde I_{E,\Omega}}{c}
=\int dE \frac{f_{E,\mu}}{(\hbar c)^3}\,\frac{E^2}{(2\pi)^2}
\end{equation}
in units of ${\rm cm}^{-3}$, a quantity which remains differential with regard to $\mu$.
Integrating it over $d\mu$ gives us the local number density $n$ of the given neutrino
species.

We define the local specific flux density (of energy) in the radial direction in the form
\begin{equation}
F_{E}=\int d\Omega\,\mu\,I_{E,\Omega}=
\int d\Omega\,\mu\,\frac{f_{E,\Omega}}{\hbar^3 c^2}\,\frac{E^3}{(2\pi)^3}
\end{equation}
in units of ${\rm cm}^{-2}~{\rm s}^{-1}$. The corresponding
energy flux density is
\begin{equation}
F=\int dE\,d\Omega\,\mu\,I_{E,\Omega}=
\int dE\,d\Omega\,\mu\,\frac{f_{E,\Omega}}{\hbar^3 c^2}\,\frac{E^3}{(2\pi)^3}
\end{equation}
in units of ${\rm MeV}~{\rm cm}^{-2}~{\rm s}^{-1}$. Finally, the luminosity
of the entire SN at a given radius $r$ is
\begin{equation}
\label{eq:luminosityint}
L=4\pi r^2 F=4\pi r^2
\int dE\,d\Omega\,\mu\,I_{E,\Omega}
\end{equation}
in units of ${\rm MeV}~{\rm s}^{-1}$. The corresponding number-quantities
(symbols with tildes) are the same expressions with $E^{-1}$ under the integrals.
In particular, $\tilde L$, the ``number luminosity,'' is the total number of
neutrinos per second of the given species streaming outward through a surface
of radius $r$.

At large distances from the SN, we usually define the average neutrino energy
in the form
\begin{equation}
\langle E\rangle = \frac{F}{\widetilde F}=
\frac{\int dE\,d\Omega\,\mu\,\frac{f_{E,\Omega}}{\hbar^3 c^2}\,\frac{E^3}{(2\pi)^3}}
{\int dE\,d\Omega\,\mu\,\frac{f_{E,\Omega}}{\hbar^3 c^2}\,\frac{E^2}{(2\pi)^3}}\,,
\label{eq:firstmom}
\end{equation}
which is the average energy of the neutrino flux at distance $r$. In deeper regions around
and below the weak decoupling region, we may also consider the
local average energy, without weighting it with the angular
projection factor $\mu$, so
\begin{equation}
\label{eq:firstmom0}
\langle E\rangle_0 =
\frac{\int dE\,d\Omega\,\frac{f_{E,\Omega}}{\hbar^3 c^2}\,\frac{E^3}{(2\pi)^3}}
{\int dE\,d\Omega\,\frac{f_{E,\Omega}}{\hbar^3 c^2}\,\frac{E^2}{(2\pi)^3}}\,.
\end{equation}
It is this quantity which we have shown in Fig.~\ref{fig:nuvelocity} as temperature ($T_0=
\langle
E\rangle_0/3.15$).

\section{Data Structure} \label{sec:nudata}

 The quantity provided by SN simulations is the monochromatic
  neutrino intensity for each flavor $\nu_\alpha$, integrated over the
  energy bin centered on $E_k$, for each radial point $r_i$ and zenith-angle
  $\mu_{ij}$.  Those data can be downloaded from the
  \href{http://wwwmpa.mpa-garching.mpg.de/ccsnarchive/index.html}{Garching
    SN Archive} upon request.
    
    We here provide data useful for
neutrino oscillation studies for three progenitors and three selected
post-bounce times as
\href{http://wwwmpa.mpa-garching.mpg.de/ccsnarchive/data/Tamborra2017/}
     {supplementary material} and  give brief instructions on how
     to read the data files. All data refer to quantities in a
     coordinate frame that moves with the matter fluid.

The files named as ``radial-neutrino-properties-time-SNmass.dat'' list the
angle-integrated neutrino emission properties  as a function of the radius
($r_i$). The first column lists the radius. The luminosities $L(r_i)$ of
$\nu_e$, $\bar{\nu}_e$ and $\nu_x$ are stored in the second, third, and
 fourth columns, respectively. The first energy moments, defined as in
Eq.~(\ref{eq:firstmom}), for $\nu_e$, $\bar{\nu}_e$ and $\nu_x$ are stored in
the fifth, sixth and seventh columns. The second energy moments for these
flavors are stored in the eighth, ninth and tenth columns. By using these data
and Eq.~(\ref{eq:alphafit})  one can reconstruct the variation of the
neutrino energy spectra in the SN comoving frame as a function of the
distance from the proto-neutron star radius.

The files named as ``angular-quantities-nualpha-time-SNmass.dat'' list the
angle- and radius-dependent neutrino emission properties for each flavor
$\nu_\alpha$ and for each selected post-bounce time and progenitor mass. The
radius $r_i$ is listed in the first column, $\mu_{ij}$ is reported in the
second column. Note that the binning in $\mu_{ij}$ is not uniform as a function
of $r_i$ because of the tangent-ray discretization of the Boltzmann transport
equation (see~\cite{RamppJanka2002} for more details). Moreover, degenerate
angular zones of measure zero may be present around $\theta=\pi/2$ due to
the peculiar angular-grid formulation.

The third column of the file ``angular-quantities-nualpha-time-SNmass.dat''
lists the differential (with respect to $\mu$, i.e., per unit interval of the
cosine of the zenith angle)
neutrino number luminosity (0-th moment, in units of s$^{-1}$)
defined as
\begin{equation}
 \tilde{L}_{\mu}(r_i,\mu_{ij}) = 4 \pi r_i^2 \sum_{k=1}^{N_E} 2 \pi \tilde{I}_{ijk}\ .
\end{equation}
We recover the neutrino number luminosity $\tilde{L}(r_i)$ 
by integrating the equation above over $\mu_{ij}$.
The fourth column represents the first energy moment (in units of MeV s$^{-1}$):
\begin{equation}
 L_{\mu}(r_i,\mu_{ij}) = 4 \pi r_i^2 \sum_{k=1}^{N_E} 2 \pi I_{ijk}\ ,
\label{eq:mom1}
\end{equation}
while the second energy moment (in units of MeV$^2$ s$^{-1}$) is reported in fifth column
and it
is defined as
\begin{equation}
 S_{\mu}(r_i,\mu_{ij}) = 4 \pi r_i^2 \sum_{k=1}^{N_E} 2 \pi I_{ijk} E_k\ .
\label{eq:mom2}
\end{equation}
The differential neutrino density  is stored in the sixth column (in units of cm$^{-3}$) and it is
defined
as
\begin{equation}
\frac{\tilde{I}_{\mu}(r_i,\mu_{ij})}{c} = 2 \pi c^{-1} \sum_{k=1}^{N_E} \tilde{I}_{ijk}\ .
\end{equation}

For all SN models, we used 21 nearly geometrically spaced energy bins up to
380 MeV and a number tangent rays and radial grid points that vary as functions
of post-bounce time: $N_\mu = 510, N_r = 235$  for  $t=61$ ms, $N_\mu = 836,
N_r = 398$ for $t = 256$~ms and $N_\mu = 838, N_r = 399$ for $t = 550$~ms for
the $11.2 M_\odot$ model;  $N_\mu = 672, N_r = 316$  for $t=150$~ms, $N_\mu =
824, N_r = 392$ for $t=280$ ms and $N_\mu = 914, N_r = 437$ for $t=500$ ms
for the $15 M_\odot$ SN progenitor;  $N_\mu = 510, N_r = 235$ for $t=63$~ms,
$N_\mu = 780, N_r = 370$ for $t = 252$~ms and $N_\mu = 784, N_r=372$ for $t =
352$~ms for the $25 M_\odot$ SN progenitor.
For all studied progenitors and post-bounce times, 
we also provide data on the matter density profile, the electron abundance,
the velocity profile, central and boundary values of the energy bins, and the
radial grid.

\bibliographystyle{apj}
\bibliography{angular}{}

\end{document}